\documentclass[12pt]{amsart}
\usepackage{amssymb,amsmath,amsthm,amscd,float,color,mathtools}
\makeatletter
\renewcommand\@biblabel[1]{#1.}
\makeatother

\theoremstyle{remark}

\numberwithin{theorem}{section} \numberwithin{equation}{section}

\newcommand{\Z}{\mathbb{Z}}

\newcommand{\im}{\textnormal{Im}\;\;\!\!\!}

\newcommand{\supp}{\textnormal{supp}}

\newcommand{\sltwoz}{\operatorname{SL}_2(\mathbb{Z})}
\newcommand{\sltwoc}{\operatorname{SL}_2(\mathbb{C})}

\newcommand{\sptwoz}{\operatorname{Sp}_4(\mathbb{Z})}
\newcommand{\sptwor}{\operatorname{Sp}_4(\mathbb{R})}
\newcommand{\sponez}{\operatorname{Sp}_2(\mathbb{Z})}
\newcommand{\spgz}{\operatorname{Sp}_{2g}(\mathbb{Z})}

\begin{document}

\title[K3 surfaces, modular forms, and non-geometric compactifications]{K3 
surfaces, modular forms, and non-geometric heterotic compactifications}

\author{Andreas Malmendier and David R. Morrison}

\date{\today}

\address{Department of Mathematics and Statistics, Colby College,
Waterville, ME 04901}
\curraddr{Department of Mathematics and Statistics, Utah State University,
Logan, UT 84322}
\email{andreas.malmendier@usu.edu}

\address{Department of Mathematics, University of California,
Santa Barbara, CA 93106} \email{drm@math.ucsb.edu}

\date{}

\begin{abstract}
We construct non-geometric compactifications 
by using the F-theory dual of the heterotic string compactified on
a two-torus, together with
a close connection between Siegel modular forms of genus two
and the equations of certain K3 surfaces.  The modular group mixes together
the K\"ahler, complex structure, and Wilson line moduli of the torus
yielding  weakly coupled heterotic string compactifications
which have no large radius interpretation.
\end{abstract}



\maketitle

\section*{Introduction}

The traditional approach to producing low-dimensional physical models
out of high-dimensional theories such as the string theories and M-theory
has been to use a specific geometric compactification of the ``extra''
dimensions and derive an effective description of the lower-dimensional
theory from the choice of geometric compactification.  However, it has
long been recognized that there are other possibilities:  for example,
one can couple perturbative string theory to an arbitrary superconformal
two-dimensional theory (geometric or not) to obtain an effective perturbative
string compactification in lower dimensions.

One way of making an analogous construction in non-perturbative
string theory is to exploit 
the nonperturbative duality transformations which relate various compactified
string theories (and M-theory) to each other.  This idea was the basis of the
construction of F-theory \cite{Vafa:1996xn}, and more recently was
used in constructions involving the type II theories \cite{Hellerman:2002ax}
and the heterotic theories \cite{nongeometries}.
We pursue a further non-geometric construction of heterotic compactifications
in this paper.

Our construction of such models relies on a very concrete relationship
between modular forms on the moduli space of
certain K3 surfaces and the equations of those
K3 surfaces \cite{MR2427457, arXiv:1004.3335, arXiv:1004.3503}.  The K3 surfaces in question have a large collection of
algebraic curve classes on them, generating a lattice known as
$\Lambda^{1,1}\oplus E_8(-1) \oplus E_7(-1)$.
The presence of these classes
restricts the form of
moduli space,\footnote{We use the term ``moduli space'' here as mathematicians
do, denoting the parameter space of the geometric objects.}
which turns out to be a space admitting  {\em modular forms.}
The modular forms in question are the Siegel modular forms of genus two,
which have previously made some appearances in the study of string
compactification.\footnote{For an early appearance, see \cite{Curio:1997si} 
and references therein.  More recently \cite{Martucci:2012jk,arXiv:1308.0553}, 
U-duality of type IIB compactifications
on K3 surfaces and the connection to modular forms of genus two
was used to construct non-geometric compactifications
analogous to the ones studied in this paper.}

The close connection between modular forms and equations allows us to
mimic the basic F-theory construction, and build
an interesting class of non-geometric heterotic compactifications
which have duals described
in terms of families of K3 surfaces.
The starting point is the heterotic string compactified on a torus,
and we exploit the non-perturbative duality symmetries which this
theory possesses.
We give the construction in considerable detail.

The paper begins with a review of F-theory in Section~\ref{sec:F}
and then proceeds to give a construction
of non-geometric heterotic compactifications in Section~\ref{sec:nongeom}.  
These compactifications
require certain $5+1$-dimensional soliton solutions
(serving as sources for scalar fields) whose heterotic construction is
discussed in Section~\ref{sec:fiveB}.

After a brief digression in Section~\ref{sec:so32} to interpret our
construction in 
the context of the $\mathfrak{so}(32)$-heterotic string, we specialize
in Section~\ref{sec:sixD}
to the case of compactifications to six dimensions.  There we find a
surprise:  although non-geometric techniques were used for the 
construction, the models we obtain are not new, but were already known
(at least in dual form).
We conclude the paper with a discussion of this surprise and its
implications.

\section{Review of F-theory} \label{sec:F}

One of the fundamental interpretations of F-theory is in terms of
the type IIB string, where it depends on three ingredients:  an 
$\sltwoz$ symmetry of the theory, a complex
scalar field $\tau$ (the axio-dilaton) with positive imaginary part (in
an appropriate normalization)  on 
which $\sltwoz$ acts by fractional linear 
transformations, and D7-branes, which serve as a source for the
multi-valuedness of $\tau$ if $\tau$ is allowed to vary.

In a standard compactification of the type IIB string, $\tau$ is a
constant and D7-branes are absent.  Vafa's idea in proposing F-theory
\cite{Vafa:1996xn} was to simultaneously
allow a variable $\tau$ and the D7-brane sources, arriving at a new
class of models in which the string coupling is never weak.

Since we cannot use the axio-dilaton $\tau$ directly in these models,
it would be natural to identify the physically relevant quantity with
$\mathbb{H}/\sltwoz$ (where $\mathbb{H}$
denotes the complex upper half-plane), but this turns out to be
slightly too simplistic.  To obtain the full range of F-theory models,
we need instead to consider some functions of $\tau$ which are
not invariant under $\sltwoz$, but rather transform
in a specific way.  A function $f(\tau)$ which satisfies
\begin{equation}
f\left(\frac{a\tau + b}{c\tau +d}\right) = (c\tau + d)^m f(\tau)
\end{equation}
for $\bigl(\begin{smallmatrix}
a&b\\c&d
\end{smallmatrix} \bigr) \in \sltwoz $
is a {\em modular form of weight $m$ for $\sltwoz$}
and such forms turn out to provide the flexibility we need for F-theory.

A simple way to write down some modular forms of even weight $m=2k$ for
$\sltwoz$ is to use what are called
{\em normalized Eisenstein series}, defined as
\begin{equation}
 E_{2k}(\tau) = \frac1{2\zeta(2k)}
\sum_{(0,0)\ne(m,n)\in \mathbb{Z}^2}
\frac1{(m\tau+n)^{2k}},
\end{equation}
where $\zeta(2k)=\sum_{n\ge1} n^{-2k}$ is Riemann's zeta function.
These normalized Eisenstein series have a Fourier expansion
(in $q=e^{2\pi i\tau}$) of the form
\begin{equation}
 E_{2k}(\tau) = 1 + O(q),
\end{equation}
which is the reason for including the normalization factor.
It is known that $E_4(\tau)$ and $E_6(\tau)$ generate the entire
ring of modular forms for $\sltwoz$.
The combination
\begin{equation}
\begin{aligned} \Delta_{12}(\tau) &= 4\left(-\frac13\, E_4(\tau)\right)^3  + 
27 \left(-\frac2{27}\, E_6(\tau)\right)^2\\
&= -\frac4{27}\, E_4(\tau)^3 + \frac4{27}\, E_6(\tau)^2
\end{aligned}
\end{equation}
plays a special role in the theory.\footnote{The convention in F-theory 
slightly differs from the one used in number theory, where the discriminant 
is identified with the 24th power of the Ramanujan tau function, 
defined to be $\eta^{24}(\tau) = q \prod_{n \ge 1} (1-q^n)^{24}$. With our 
conventions for $\Delta_{12}$, we have 
$\Delta_{12}(\tau)  = - 2^8 \, \eta^{24}(\tau)$.}
In particular, if we compactify the parameter space 
$\sltwoz$ to $\overline{\sltwoz}$,
then $\Delta_{12}(\tau)$ extends to the compactification and
vanishes on the boundary (which corresponds to the $q\to0$ limit).

An F-theory compactification (regarded as a compactification of the type IIB
string with variable axio-dilaton) takes as its
staring point a compact space $W$, a complex 
line bundle $\mathcal{L}$ on $W$ and sections $f(w)$ and $g(w)$ of the 
associated
bundles $\mathcal{L}^{\otimes 4}$ and $\mathcal{L}^{\otimes 6}$.
Then there is a (possibly non-supersymmetric)
 F-theory model 
with a variable $\tau$ function and $\sltwoz$
symmetry,
in which $f(w)$ is identified with $-\frac13E_4(\tau)$ and $g(w)$ is identified
with $-\frac2{27}E_6(\tau)$.  
One must also insert seven-branes of various kinds along the zeros of
\begin{equation}
 \Delta(w):=4f(w)^3+27g(w)^2.
\end{equation}
The geometry behind this construction is a beautiful story from 19th-century
mathematics: the Weierstrass $\wp$-function.  In order to define
a doubly-periodic meromorphic function in the complex plane
(with periods $1$ and $\tau$), Weierstrass
introduced a function of $z\in\mathbb{C}$ and $\tau$:
\begin{equation}
 \wp(z,\tau) = \frac1{z^2} + \sum_{(m,n)\ne(0,0)} \left(
\frac1{(z-m\tau-n)^2} - \frac1{(m\tau+n)^2}\right).
\end{equation}
This has a Laurent expansion (using the normalized Eisenstein series
as well as the special values $\zeta(4)=\pi^4/90$ and $\zeta(6)=\pi^6/945$):
\begin{equation}
\begin{aligned} \wp(z,\tau) &= z^{-2}+ 6\, \zeta(4) \, E_4(\tau) \, z^2 + 10\,
 \zeta(6)\, E_6(\tau) \, z^4
+O(z^6),\\
&= z^{-2}+ \frac{\pi^4}{15} \, E_4(\tau)\, z^2 + \frac{2\pi^6}{189}\, 
E_6(\tau)\, z^4
+O(z^6),
\end{aligned}
\end{equation}
from which Laurent expansions for $(\wp(z,\tau)')^2$ and $\wp(z,\tau)^3$
can be calculated:
\begin{align} (\wp(z,\tau)')^2 &= 4z^{-6}-\frac{8\pi^4}{15} \, E_4(\tau) \, z^{-2}-\frac{32\pi^6}{189} \, E_6(\tau)+O(z)\;,\\
 \wp(z,\tau)^3 &= z^{-6} + \frac{\pi^4}5\, E_4(\tau)\, z^{-1}+\frac{6\pi^6}{189} \, E_6(\tau)+O(z).
\end{align}
It follows that
\begin{equation}
 (\wp(z,\tau)')^2 - 4\, \wp(z,\tau)^3 +\frac{4\pi^4}3E_4(\tau)\, \wp(z,\tau)=
-\frac{8\pi^6}{27}E_6(\tau)+O(z)=
-\frac{8\pi^6}{27}E_6(\tau).
\end{equation}
(This is an exact expression since the left hand side is an
entire holomorphic function which is bounded since it is doubly-periodic,
and hence constant.)
If we set $x:=\frac1{\pi^2}\wp(z,\tau)$ and $y:=\frac1{2\pi^3}\wp(z,\tau)'$, we
find an equation for
the elliptic curve with modular parameter $\tau$:
\begin{equation}
 y^2 = x^3 - \frac13\, E_4(\tau)\, x - \frac2{27}\, E_6(\tau).
\end{equation}

Conversely, if we start from an elliptic curve with an equation of
the form
\begin{equation}\label{eq:ell}
 y^2=x^3+fx+g
\end{equation}
which is nonsingular,
then we can recover $\tau$ up to $\sltwoz$ transformation as
\begin{equation}
 \tau = \frac{\int_{\gamma_2} \frac{dx}{\sqrt{x^3+fx+g}}}
   {\int_{\gamma_1}\frac{dx}{\sqrt{x^3+fx+g}}} ,
\end{equation}
where $(\gamma_1,\gamma_2)$ is an oriented basis of the first homology
of the elliptic curve, such that $f = -\frac{\lambda^4}3 E_4(\tau)$,
$g=-\frac{2\lambda^6}{27}E_6(\tau)$ for some nonvanishing scale factor $\lambda$.
The condition for nonsingularity of \eqref{eq:ell}
is that the quantity
\begin{equation} \Delta:=4f^3+27g^3
\end{equation}
does not vanish.
It is this close connection between geometry and modular forms which allows
the construction of families of elliptic curves with certain knowledge of
the behavior of $\tau$ in such families.

To understand when the corresponding F-theory models are supersymmetric,
we follow the duality between F-theory and M-theory.  That duality is
based on a key fact:  Compactifying M-theory on a torus whose complex
structure is labeled by $\tau$ and whose area is $A$ gives a model dual
to the type IIB string compactified on a circle of radius\footnote{The radius
is measured in Einstein frame.} $A^{-3/4}$
whose axio-dilaton has value $\tau$ \cite{Schwarz:1995dk,Aspinwall:1995fw}.  
This then gives a connection
between the F-theory construction and a dual geometric compactification
of M-theory:  if the F-theory model is further compactified on $S^1$ 
(which can be done without breaking any supersymmetry that might be present), 
a model will be obtained which is dual to M-theory compactified on
the total space of the family
\begin{equation} y^2=x^3+f(w)\, x+g(w) \end{equation}
of elliptic curves over $W$.  (The insertions of seven-branes in the F-theory
model go over to singular elliptic fibers in the M-theory model
which may require special treatment, but
for generic sections $f$ and $g$ the total space of the family is nonsingular
and the compactification makes sense as it stands.)

One can then ask whether the geometric M-theory model breaks or preserves
supersymmetry, and the answer is known: 
supersymmetry is preserved exactly when the total space of
the family is a Calabi--Yau manifold, which happens exactly when
the line bundle $\mathcal{L}$ is the anti-canonical bundle of
the base, i.e., $\mathcal{L}=\mathcal{O}_W(-K_W)$.  In this way,
 we recover the familiar conditions
for a supersymmetric F-theory compactification.

To complete the story, we need to know what types of seven-branes need to
be inserted.  The answer here comes from algebraic geometry, through
work of Kodaira
 \cite{MR0184257} and N\'eron \cite{MR0179172} which
classifies the possible singular limits in one-parameter families
of elliptic curves and thereby gives a catalog of the different types of 
seven-branes
which must be inserted.  This catalog is by now well-known, but we will 
reproduce it in Table~\ref{tab:kodaira}, in which labels for
the types of seven-branes are given using
 Kodaira's notation.  The type of brane depends on the orders of vanishing
of $f$, $g$, and $\Delta$ at the singular point $P$, and determines both the
type of singularity which appears in the M-theory dual, and the transformation
in $\sltwoz$ which describes how $\tau$ changes when the singular point
is encircled.  
The $I_n$ type corresponds to a stack of $n$ D7-branes, while the $I_n^*$
type corresponds to a stack of $n{+}4$ D7-branes on top of an orientifold plane.
The last line of the table (labeled ``non-minimal'') can
be avoided by a suitable choice of line bundle for the Weierstrass model.

\begin{table}
{\footnotesize
\begin{center}
\begin{tabular}{|c|c|c|c|c|c|} \hline
brane type&$\operatorname{ord}_P(f)$&$\operatorname{ord}_P(g)$
&$\operatorname{ord}_P(\Delta)$
&singularity&transformation\\ \hline\hline
$I_0$&$\ge0$&$\ge0$&$0$&none
&$\begin{pmatrix}\hphantom{-}1&\hphantom{-}0\\\hphantom{-}0&\hphantom{-}1\end{pmatrix}$\\ \hline
$I_n$, $n\ge1$&$0$&$0$&$n$&$A_{n-1}$
&$\begin{pmatrix}\hphantom{-}1&\hphantom{-}n\\\hphantom{-}0&\hphantom{-}1\end{pmatrix}$\\ \hline
$II$&$\ge1 $&$   1  $&$    2 $&  none
&$\begin{pmatrix}\hphantom{-}1&\hphantom{-}1\\-1&\hphantom{-}0\end{pmatrix}$\\ \hline
$III$&$  1 $&$   \ge2 $&$   3 $&$  A_1$
&$\begin{pmatrix}\hphantom{-}0&\hphantom{-}1\\-1&\hphantom{-}0\end{pmatrix}$\\ \hline
$IV$&$ \ge2 $&$  2  $&$    4 $&$  A_2$
&$\begin{pmatrix}\hphantom{-}0&\hphantom{-}1\\-1&-1\end{pmatrix}$\\ \hline
$I_0^*$&$\ge2$&$\ge3$&$6$&$D_{4}$
&$\begin{pmatrix}-1&\hphantom{-}0\\\hphantom{-}0&-1\end{pmatrix}$\\ \hline
$I_n^*$, $n\ge1$&$2$&$3$&$n+6$&$D_{n+4}$
&$\begin{pmatrix}-1&-n\\\hphantom{-}0&-1\end{pmatrix}$\\ \hline
$IV^*$&$\ge3$&$  4$  &$  8$&$   E_6$
&$\begin{pmatrix}-1&-1\\\hphantom{-}1&\hphantom{-}0\end{pmatrix}$\\ \hline
$III^*$&$  3 $&$   \ge5 $&$   9 $&$  E_7$
&$\begin{pmatrix}\hphantom{-}0&-1\\\hphantom{-}1&\hphantom{-}0\end{pmatrix}$\\ \hline
$II^*$&$ \ge4$&$   5   $&$   10 $&$  E_8$
&$\begin{pmatrix}\hphantom{-}0&-1\\\hphantom{-}1&\hphantom{-}1\end{pmatrix}$\\ \hline
non-minimal&$\ge4$&$\ge6$&$\ge12$&non-canonical&--\\ \hline
\end{tabular}
\end{center}
\medskip
\caption{Kodaira--N\'eron classification of singular fibers and monodromy}\label{tab:kodaira}
}
\end{table}

\section{Non-geometric heterotic models}
\label{sec:nongeom}

By analogy, we now wish to study as our basic theory the heterotic string
compactified on $T^2$ to produce an eight-dimensional effective theory.
(This will be our analogue of the type IIB string in the previous section.)
This effective theory has a complex scalar field which, after symmetries
are taken into account, takes its values in the Narain space%
\footnote{To better match both the algebraic geometry and modular
forms literature, we take the Narain lattices for compactifications of the 
heterotic string on a $d$-torus to have signature $(d,16+d)$, which is the
opposite of the usual convention in string theory.}
\cite{Narain:1985jj}
\begin{equation}
\mathcal{D}_{2,18}/O(\Lambda^{2,18})
\end{equation}
which is the quotient
 of the symmetric space for $O(2,18)$,
\begin{equation}
\mathcal{D}_{2,18} := (O(2)\times O(18))\backslash O(2,18) ,
\end{equation}
by the automorphism group $O(\Lambda^{2,18})$ 
of the unique integral even unimodular lattice $\Lambda^{2,18}$ of
signature
 $(2,18)$.  (This discrete group is sometimes called $O(2,18,\mathbb{Z})$.)
In an appropriate limit, this space decomposes as a product of 
spaces parameterizing the K\"ahler and complex structures on $T^2$
as well as Wilson line expectation values around the two generators of $\pi_1(T^2)$ 
(see \cite{Narain:1986am}).  However,
that decomposition is only preserved by a parabolic subgroup $\Gamma\subset 
O(\Lambda^{2,18})$, which is much smaller.  
Letting the moduli of the entire space vary 
arbitrarily (i.e., employing the full $O(\Lambda^{2,18})$ symmetry) 
will produce a compactification
which has a right to be called non-geometric, because the K\"ahler and 
complex structures on $T^2$, and the Wilson line values, are not distinguished
under the $O(\Lambda^{2,18})$-equivalences but instead are mingled together.
With no K\"ahler class, we lose track of geometry.\footnote{Note that the
heterotic dilaton is {\em not}\/ affected by this group action, so weakly
coupled models of these non-geometric heterotic strings will exist.}

The construction we will give of non-geometric heterotic compactifications
actually uses an index $2$ subgroup $O^+(\Lambda^{2,18})
\subset O(\Lambda^{2,18})$ defined by
\begin{equation} 
O^+(\Lambda^{2,18}):= O^+(2,18) \cap O(\Lambda^{2,18}) ,
\end{equation}
where $O^+(p,q)$ denotes the subgroup of $O(p,q)$
preserving the orientation on positive
$p$-planes.  The group $O^+(\Lambda^{2,18})$ 
is the maximum subgroup of $O(\Lambda^{2,18})$ whose action
preserves the complex structure on the symmetric space, and thus is the
maximal subgroup for which modular forms can be holomorphic.  The corresponding
quotient
\begin{equation} \label{k3moduli}
\mathcal{D}_{2,18}/ O^+(\Lambda^{2,18})
\end{equation}
is a degree two cover of the Narain moduli space.  The group 
$O^+(\Lambda^{2,18})$ is still large enough to thoroughly mix the K\"ahler, 
complex, and Wilson line moduli.

The quotient space \eqref{k3moduli} is the parameter space for
elliptically fibered K3 surfaces
with a section:
this is a statement of the duality between F-theory and the heterotic string
in eight dimensions \cite{Vafa:1996xn}, 
and the identification of the discrete group for this moduli problem
as $O^+(\Lambda^{2,18})$
is well-known in the mathematics literature (see, for example,
\cite{MR2336040}).\footnote{The slight mismatch in duality groups between
$O(\Lambda^{d,16+d})$ for the heterotic string compactified on $T^d$
and $O^+(\Lambda^{d,16+d})$
for the dual theory occurs for the type I' dual (interpreted as a real
K3 surface) when $d=1$ \cite{Cachazo:2000ey},
the F-theory dual when $d=2$ as indicated here, 
the M-theory dual when $d=3$ (described
in \cite{stringK3} and 
based on \cite{MR849050,MR1102278,MR1066174}), 
and the type IIA dual when $d=4$ \cite{Nahm:1999ps}.  
It would be interesting to have a more complete understanding of
this mismatch.}
To use these K3 surfaces in a similar way
to the way that elliptic curves were used in constructing F-theory,
we would need a close connection between
$O^+(\Lambda^{2,18})$-modular
forms and the equations of
the corresponding elliptically fibered K3 surfaces; unfortunately
such a connection is not known.

However, by making a simple and natural restriction on our heterotic theories,
we can find and
exploit such a connection.  Namely, let us consider heterotic models
with only a single nonzero Wilson line expectation value.  For definiteness, we
restrict to the $\mathfrak{e}_8\oplus\mathfrak{e}_8$ heterotic string,
and note that asking for an unbroken gauge algebra of
$\mathfrak{e}_8\oplus\mathfrak{e}_7$ will ensure that only a single
Wilson line expectation value is nonzero.  (There is a similar story for the
$\mathfrak{so}(32)$ string which we will describe in Section~\ref{sec:so32}.)

Let $L^{2,3}$ be the lattice of signature $(2,3)$ which is
the orthogonal complement of $E_8(-1) \oplus E_7(-1)$
in $\Lambda^{2,18}$.
By insisting that the Wilson lines values associated to the\footnote{The sign change on $E_8$ and $E_7$
is due to our sign conventions about the Narain lattice.}
$E_8(-1)\oplus E_7(-1)$ sublattice
be trivial (which leaves the algebra $\mathfrak{e}_8\oplus\mathfrak{e}_7$
unbroken), we restrict to those heterotic vacua
parameterized by the space
\begin{equation}
\mathcal{D}_{2,3}/O(L^{2,3}).
\end{equation}
The corresponding degree two cover is
\begin{equation} \mathcal{D}_{2,3}/O^+(L^{2,3}),
\end{equation}
and this space parameterizes elliptically fibered K3 surfaces with section which
have one fiber 
of Kodaira type $III^*$ or worse
and another fiber
of Kodaira type precisely\footnote{As we will see in Section~\ref{sec:sixD},
assuming that this fiber is {\em precisely}\/ type $II^*$ avoids ``pointlike
instantons'' on the heterotic dual after compactification to dimension
six or below, at least for general moduli.} $II^*$. 
Such K3 surfaces contain the lattice $\Lambda^{1,1} \oplus E_8(-1)\oplus E_7(-1)$
inside their N\'eron--Severi lattice, and are often referred to as ``lattice-polarized K3 surfaces''.  
The $\Lambda^{1,1}$ summand is generated by the classes
of the fiber and the section of the elliptic fibration.\footnote{For the
lattice embedding of $\Lambda^{1,1}$ to correspond to an elliptic fibration with section we also need to require
that its image in the N\'eron--Severi lattice contains a pseudo-ample class.}

As we will describe below, the modular forms for 
$O^+(L^{2,3})$ have the desired property:  there is a close geometric
connection to the corresponding lattice-polarized K3 surfaces.
(A similar picture was developed in earlier work in the case of
no nontrivial Wilson line expectation values,
using modular forms for $O^+(\Lambda^{2,2})$
\cite{nongeometries}.)

Let $\mathbb{H}_g$ denote the Siegel upper half-space of genus $g$, on
which the Siegel modular group $\spgz$ acts.
As explained in Appendix~\ref{app:discrete},  for $g=2$ there is a homomorphism
$\sptwor \to O^+(2,3)$ which induces an isomorphism
\begin{equation} \label{eq:h2iso}
\mathbb{H}_2 \cong \mathcal{D}_{2,3}.
\end{equation}
By a result of
Vinberg \cite{vinberg-siegel} reviewed in Appendix~\ref{app:discrete},
the image of $\sptwoz \to O^+(L^{2,3})$ is a subgroup of index
$2$,
and the ring of $O^+(L^{2,3})$-modular forms turns out to correspond
to the ring of Siegel modular forms with $g=2$ 
{\it of even weight.}\footnote{Note that because the groups are different,
families of genus two curves (such as were used in 
\cite{Martucci:2012jk,arXiv:1308.0553}) are not equivalent to the families
of elliptic K3 surfaces needed for our construction.}
Igusa \cite{MR0141643} showed that this latter ring is a polynomial ring in
four free generators of degrees $4$, $6$, $10$ and $12$.  We explicitly
describe Igusa's generators $\psi_4$, $\psi_6$, $\chi_{10}$
and $\chi_{12}$ in Section~\ref{Siegel_modular_forms}.  (Igusa
later showed \cite{MR0229643} that for the full
ring of $\sptwoz$-modular forms, one needs an additional generator
$\chi_{35}$, also described in Section~\ref{Siegel_modular_forms}, 
which is algebraically dependent on the others.  In fact,
$\chi_{35}^2$ is an an explicit polynomial in  $\psi_4$, $\psi_6$,
$\chi_{10}$ and $\chi_{12}$ which is given in \eqref{chi35sqr}.)

The key geometric fact, due in different forms to Kumar \cite{MR2427457}
and to Clingher--Doran \cite{arXiv:1004.3335, arXiv:1004.3503}, is the equation for an elliptically fibered
K3 surface whose periods give the point $\underline{\tau}$ in the Siegel
upper halfspace $\mathbb{H}_2$, 
with the coefficients in the equation being Siegel modular
forms of even weight.  (This is analogous to the  Weierstrass equation
for the elliptic curve $\mathbb{C}/\langle 1,\tau\rangle$
with coefficients being Eisenstein series in $\tau$).  That equation is:
\begin{equation}\label{eq:imp}
 y^2 = x^3 - t^3 \, \left( \frac{1}{48} \psi_4(\underline{\tau}) \, t +4\chi_{10}(\underline{\tau})
\right) \, x + t^5 \, \left(  t^2 -  \frac{1}{864} \psi_6(\underline{\tau}) \,
t+\chi_{12}(\underline{\tau})\right).
\end{equation}
Just as in the elliptic curve case, the statement has two parts:  
starting from $\underline{\tau}$, we obtain the equation of a K3 surface
\eqref{eq:imp}.  But conversely, if we start with a K3 surface $S$ with an
equation of the form\footnote{
The reader may wonder why the coefficients of $x^3$, $y^2$, and $t^7$
have been set equal to $1$ in \eqref{eq:genform}.  The choice of
coefficient $1$ for $x^3$ and $y^2$ is a familiar one, and derives from
an analysis by Deligne \cite{MR0387292} of families of elliptic curves: assuming that
all fibers are generalized elliptic curves of an appropriate kind, it
follows that the coefficients of $x^3$ and $y^2$ never vanish, and then
by a change of coordinates these coefficients can be set to $1$.
The story for $t^7$ is similar: we are assuming that the Kodaira fiber
at $t=\infty$ is exactly $II^*$, and this implies that the $t^7$ term
in the equation must always be present.  Thus, we could allow a coefficient
$\alpha$ for $t^7$ but it would never be allowed to vanish; as a consequence,
the change of coordinates $(x,y,t)\mapsto (x/\alpha^2,y/\alpha^3,t/\alpha)$
would be well-defined, and would map
$x^3-y^2+\alpha t^7$ to $\alpha^{-6}(x^3-y^2+t^7)$.  In other
words, by making such a change of coordinates and then rescaling the
entire equation by $\alpha^6$, we may assume that the coefficient of
$t^7$ is $1$.
}
\begin{equation} \label{eq:genform}
y^2 = x^3 + a \, t^4 \, x + b \, t^6 + c\, t^3 \, x + d\, t^5 + t^7,
\end{equation}
and we determine a point in 
$\mathcal{D}_{2,3}$
by calculating the periods of the holomorphic $2$-form on $S$
over a basis of the orthogonal complement of 
$\Lambda^{1,1} \oplus E_8(-1)\oplus E_7(-1)$ in $H^2(S,\mathbb{Z})$
(which in turn determines $\underline{\tau}\in \mathbb{H}_2$
using the isomorphism \eqref{eq:h2iso}), then for some nonzero scale factor
$\lambda$,
\begin{equation}
a = -\frac{\lambda^4}{48}\, \psi_4(\underline{\tau}), \
b = -\frac{\lambda^6}{864}\, \psi_6(\underline{\tau}), \
c = -4\, \lambda^{10}\, \chi_{10}(\underline{\tau}), \
d = \lambda^{12}\, \chi_{12}(\underline{\tau}).
\end{equation}
We verify in Appendix~\ref{K3fibration} that the K3 surface defined by
\eqref{eq:imp} agrees with
the ones found by Kumar and by Clingher--Doran.

The strategy for constructing a non-geometric heterotic compactification
is now clear.  Start with  compact manifold $Z$ and a line bundle
$\Lambda$ on $Z$.  Pick sections $a(z)$, $b(z)$, $c(z)$, and $d(z)$ of
$\Lambda^{\otimes 4}$, $\Lambda^{\otimes 6}$, $\Lambda^{\otimes 10}$, and
$\Lambda^{\otimes 12}$, respectively.
Then there is a non-geometric heterotic compactification on $Z$ with variable
$\underline{\tau}$ and $O^+(L^{2,3})$ symmetry for which 
\begin{equation}
\begin{aligned}
a(z) &= -\frac{1}{48}\, \psi_4(\underline{\tau}), \\
b(z) &= -\frac{1}{864}\, \psi_6(\underline{\tau}), \\[0.4em]
c(z) &= -4 \, \chi_{10}(\underline{\tau}), \\[0.6em]
d(z) &= \chi_{12}(\underline{\tau}).
\end{aligned}
\end{equation}
(We can eliminate the scale factor $\lambda$, if any, by making a change
of coordinates $(x,y,t)\mapsto (\lambda^{14}x,\lambda^{21}y,\lambda^6t)$.)
Appropriate five-branes must be inserted on $Z$ as dictated by the geometry
of the corresponding family of K3 surfaces
\begin{equation} \label{eq:family}
y^2 = x^3 + a(z)\,t^4x + b(z)\,t^6 + c(z)\,t^3x + d(z)\,t^5 + t^7.
\end{equation}
We will explore these five-branes in the next section.

To understand when the non-geometric heterotic compactifications
we have constructed 
are supersymmetric, we follow the duality between the heterotic string
and F-theory.  The heterotic compactification on $T^2$ with parameter
$\underline{\tau} \in \mathbb{H}_2$ is dual to the F-theory compactification
on the elliptically fibered K3 surface $S_{\underline{\tau}}$
defined by \eqref{eq:imp}, where
$t$ is a local coordinate on the base $\mathbb{P}^1$ of the elliptic fibration.
Note that at $t=\infty$, $S_{\underline{\tau}}$ has a Kodaira fiber of type
precisely $II^*$:  it can be no worse because the coefficient of $t^7$
in $\eqref{eq:imp}$ is $1$.  At $t=0$, there is a Kodaira fiber of type $III^*$
or worse.

When $a(z)$, $b(z)$, $c(z)$, and $d(z)$ are sections of line bundles over
$Z$, we wish to determine whether F-theory compactified on the elliptically 
fibered manifold \eqref{eq:family} is supersymmetric, and this in turn 
depends on whether
the total space defined by \eqref{eq:family} is itself  a 
Calabi--Yau manifold.  
The base of the elliptic fibration on the total space is a $\mathbb{P}^1$-bundle
$\pi:W \to Z$ which takes the form $W=\mathbb{P}(\mathcal{O}\oplus \mathcal{M})$
for some line bundle $\mathcal{M}$ that coincides with the normal bundle
of $\Sigma_0:=\{t=0\}$ in $W$.  Restricting the various terms in 
\eqref{eq:family} to $\Sigma_0$, we find relations
$\Lambda^{\otimes 4} \otimes \mathcal{M}^{\otimes 4} 
= \Lambda^{\otimes 10} \otimes \mathcal{M}^{\otimes 3}
= (\mathcal{L}|_{\Sigma_0})^{\otimes 4}$
and 
$
\mathcal{M}^{\otimes 7} 
=\Lambda^{\otimes 6} \otimes \mathcal{M}^{\otimes 6}
=\Lambda^{\otimes 12} \otimes \mathcal{M}^{\otimes 5}
= (\mathcal{L}|_{\Sigma_0})^{\otimes 6}$.
It follows that 
that $\mathcal{M}=\Lambda^{\otimes 6}$ and 
$\mathcal{L}|_{\Sigma_0}=\Lambda^{\otimes 7}$ (up to torsion).
In other words, our $\mathbb{P}^1$-bundle must take the form
$W=\mathbb{P}(\mathcal{O}\oplus \Lambda^{\otimes 6})$.
This property can be traced back to the fact that the coefficient
of $t^7$ in \eqref{eq:family} is $1$.

Now to check the condition for supersymmetry, note that 
\begin{equation}
-K_W = \Sigma_0 + \Sigma_\infty + \pi^{-1}(-K_Z),
\end{equation}
where $\Sigma_\infty:=\{t=\infty\}$.  Since $\Sigma_0$ and $\Sigma_\infty$
are disjoint, it follows that the condition for supersymmetry
$\mathcal{L}=\mathcal{O}_W(-K_W)$ is equivalent to
$\Lambda = \mathcal{O}_Z(-K_Z)$.

Let us briefly comment on the relationship of our construction with
the appearance of Siegel modular forms in string compactifications
involving the ``STUV'' model, as described in \cite{Curio:1997si}
and the references therein.  If we take the mirror of our family
of lattice-polarized K3 surfaces
\cite{stringK3,MR1420220}, we will obtain a family of K3 surfaces
depending on $17$ complex parameters
whose quantum K\"ahler moduli space is $\mathcal{D}_{2,3}/O^+(L^{2,3})$.
The K3 surfaces in the
new family all contain a lattice $L^{1,2}$ within $H^{1,1}$ with
the property that $L^{2,3}\cong \Lambda^{1,1}\oplus L^{1,2}$.
If $X$ is a Calabi--Yau threefold which has a one-parameter family
of such K3 surfaces on it, then the $(1,1)$ classes on $X$
also include the lattice $L^{1,2}$. Type IIA string theory compactified
on $X$ is the dual theory of the heterotic STUV model, as discussed
in \cite{Curio:1997si} and elsewhere.  It is natural that quantum
corrections of this gravitational theory would respect the symmetry group
$O^+(L^{2,3})$ and so would turn out to be related
to Siegel modular forms as well.

\section{Five-branes} \label{sec:fiveB}

The base $W$ of an elliptic fibration maps naturally to the compactification
$\overline{\mathbb{H}/\sltwoz}$ of the parameter
space $\mathbb{H}/\sltwoz$, and if this map is 
nonconstant, there must be singular fibers (at which the $j$-invariant
approaches $\infty$).  In fact, for a generic elliptic fibration, all
seven-branes will have
$j\to\infty$, and those correspond to familiar
seven-brane constructions in type IIB string theory (D7-branes, possibly
combined with orientifold planes).

The situation for fibrations of lattice-polarized K3 surfaces is very
different.  There is a Satake-Baily-Borel compactification 
\cite{MR0118775,BailyBorel} $\overline{\mathcal{D}_{2,3}/O^+(L^{2,3})}$
of the parameter space
whose boundary has codimension two, and this implies that a one-parameter
family of lattice-polarized K3 surfaces {\em need not reach the boundary!}
That would suggest that it might be possible to have a family which never
degenerates (i.e., with no brane insertions needed), 
but this is not the case:  there are codimension one loci
where some elements of $O^+(L^{2,3})$ have fixed points, and there must
always be branes associated with these fixed loci.

To find group elements with fixed points, note that
a reflection in a lattice element of square $-2$ has a fixed locus of
codimension one, belongs to
$O^+(L^{2,3})$, and does not belong to $SO^+(L^{2,3})\cong \sptwoz$.  As
a consequence, such reflections must act as $-1$ on the $\sptwoz$-modular
forms of odd weight, and the fixed locus of any such reflection must
be contained in the vanishing locus of any $\sptwoz$-modular form of odd
weight.  The modular forms of odd weight are generated by Igusa's form
$\chi_{35}$, so that form must vanish along the fixed loci of our reflections.

From the point of view of K3 geometry, if the periods are preserved by
the reflection in $\delta$ with $\delta^2=-2$, then $\delta$ must belong
to the N\'eron-Severi lattice of the K3 surface.  That is, the
lattice $\Lambda^{1,1}\oplus E_8(-1)\oplus E_7(-1)$ must be enlarged
by adjoining $\delta$.  It is not hard to show (using methods of
\cite{ISBF}, for example), that there are only two ways this enlargement
can happen (if we have adjoined a single element only):  either the lattice is extended to $\Lambda^{1,1}\oplus E_8(-1)\oplus E_8(-1)$
or it is extended to $\Lambda^{1,1}\oplus E_8(-1)\oplus E_7(-1) \oplus
\langle -2 \rangle$.  In the former case, the fibers in the elliptic fibration
become $II^*$, $II^*$ and 4 $I_1$, whereas in the latter case, the fibers
become $II^*$, $III^*$, $I_2$ and 4 $I_1$.

If we start with an elliptically fibered K3 surface \eqref{eq:genform},
then it is easy to see what the condition is for the first enhancement:
we want the fiber at $t=0$ to go from type $III^*$ to type $II^*$, and
this is achieved by setting $c=0$.

To see the second enhancement requires a computation.
Starting with \eqref{eq:genform}, we compute the discriminant of the
elliptic fibration to be
\begin{equation}
\Delta 
=
t^9 \left( 4(at+c)^3+27t(t^2+bt+d)^2 \right) 
\end{equation}
The zeros of $\Delta/t^9$ represent the location of the $I_1$ fibers,
so to find out when they coincide, we calculate the discriminant of
that polynomial of degree $5$ in $t$ (which will vanish precisely when
there are multiple roots).  That discriminant turns out to take the form
\begin{equation}
2^83^{12}\ell(a,b,c,d)^3q(a,b,c,d),
\end{equation}
where, if we assign weights $4$, $6$, $10$, $12$ to $a$, $b$, $c$, $d$,
respectively, then $\ell$ is the polynomial of weighted degree $20$
\begin{equation}\label{ell}
\ell(a,b,c,d) := a^2 \, d-a\,b\,c+c^2  \;,
\end{equation}
and $q$ is the polynomial of weighted degree $60$
\begin{equation}
\label{q-eqn}
\begin{split}
q(a,b,c,d) &:=
11664\,{d}^{5}
+864\,{a}^{3}{d}^{4}
-5832\,{b}^{2}{d}^{4}
+16\,{a}^{6}{d}^{3}
+216\,{a}^{3}{b}^{2}{d}^{3}
\\ &
-2592\,{a}^{2}bc{d}^{3}
+16200\,a{c}^{2}{d}^{3}
+729\,{b}^{4}{d}^{3}
+888\,{a}^{4}{c}^{2}{d}^{2}
-5670\,a{b}^{2}{c}^{2}{d}^{2}
\\ &
-13500\,b{c}^{3}{d}^{2}
+16\,{a}^{7}{c}^{2}d
+216\,{a}^{4}{b}^{2}{c}^{2}d
-3420\,{a}^{3}b{c}^{3}d
+4125\,{a}^{2}{c}^{4}d
\\ &
+729\,a{b}^{4}{c}^{2}d
+6075\,{b}^{3}{c}^{3}d
-16\,{a}^{6}b{c}^{3}
+16\,{a}^{5}{c}^{4}
-216\,{a}^{3}{b}^{3}{c}^{3}
\\ &
+2700\,{a}^{2}{b}^{2}{c}^{4}
-5625\,ab{c}^{5}
-729\,{b}^{5}{c}^{3}
+3125\,{c}^{6}
\end{split}
\end{equation}
(which we computed directly using computer algebra).

The role of the polynomial $\ell$ is easy to see:  it vanishes on precisely
those K3 surfaces for which $f$ and $g$ have a common zero (at $t=-c/a$).
Those are cases in which two $I_1$'s are replaced by a fiber of type $II$,
but such a change does not affect the lattice or the gauge algebra
and so these are not the K3 surfaces we are looking for.

Since the two lattice enhancements occur at $c=0$ and $q(a,b,c,d)=0$,
we predict that $c\cdot q(a,b,c,d)$ should vanish along the locus where
there is some degeneration.  Indeed it turns out (as verified in 
Section~\ref{Siegel_modular_forms}) that
\begin{equation}
q\left(-\frac1{48}\psi_4,-\frac1{864}\psi_6,-4 \, \chi_{10}, \, \chi_{12}\right)
= 2^{-8} \, \chi_{35}^2/\chi_{10},
\end{equation}
confirming the prediction.

Thus, a generic non-geometric compactification constructed from these
lattice-polarized K3 surfaces will have two types of five-branes, corresponding
to\footnote{The loci $\{c=0\}$ and $\{q(a,b,c,d)=0\}$ correspond to the
well-studied Humbert surfaces $H_1$ and $H_4$ described in 
Appendix~\ref{app:B}.} 
$c=0$ and $q(a,b,c,d)=0$.  From the heterotic side, these five-brane
solitons are easy to see.  When $q(a,b,c,d)=0$, we have an additional
gauge symmetry enhancement to include $\mathfrak{su}(2)$, and the  parameters
of the theory include a Coulomb branch for that gauge theory on which
the Weyl group $W_{\mathfrak{su}(2)}=\mathbb{Z}_2$ acts.  There is thus a five-brane
solution
in which the field has a $\mathbb{Z}_2$ ambiguity encircling the location
in the moduli space of enhanced gauge symmetry.

The other five-brane solution is similar:  at $c=0$, there is
an enhancement from
$\mathfrak{e}_7$ to $\mathfrak{e}_8$ gauge symmetry, and a similar $\mathbb{Z}_2$ acts
on the moduli space, leading to a solution with a $\mathbb{Z}_2$ ambiguity.
These two brane solutions are the analogue of the simplest brane (a single
D7-brane) in F-theory.

Finding a complete catalog of five-brane solutions for this theory
is quite challenging.
As we explain in Appendix~\ref{app:modulispaces}, the parameter space
$\mathcal{D}_{2,3}/O^+(L^{2,3})$ for our construction is closely related
to some other moduli spaces:  the moduli space
of homogeneous sextics in two variables,
the moduli space of Abelian surfaces, and the moduli space
of curves of genus two.  A version of Kodaira's
classification was given for curves of genus two by Namikawa and Ueno
\cite{MR0369362}
and this can in principle be used  to give a classification of degenerations
of these lattice-polarized K3 surfaces.  We illustrate how this works
in a number of interesting cases in Appendix~\ref{degs_and_branes}.

\section{The 
$\mathfrak{so}(32)$  heterotic string}
\label{sec:so32}

It turns out that the total space of a lattice-polarized K3 surface of
the form \eqref{eq:genform} with lattice polarization by
$\Lambda^{1,1} \oplus E_8(-1) \oplus E_7(-1)$ always
admits a second elliptic fibration with a different polarization
\cite{arXiv:1004.3503} (see also \cite{nongeometries}),
which can be related to the $\mathfrak{so}(32)$  heterotic string.
To see this, consider the birational transformation
\begin{equation}
x=X^2T, \quad y=X^2Y, \quad t=X
\end{equation}
applied to \eqref{eq:genform}.  (In applying the transformation, we
make the substitution and then divide by the common factor of $X^4$.)
The result is the equation
\begin{equation} \label{eq:so32}
\begin{aligned}
Y^2 &= X^2T^3 + aX^2T + bX^2 + cXT + dX + X^3 \\
&= X^3 + (T^3+aT + b)X^2 + (cT+d)X.
\end{aligned}
\end{equation}
To more easily see the structure, we introduce homogeneous coordinates
$[S,T]$ on the base $\mathbb{P}^1$ and write the equation as
\begin{equation} \label{eq:homog}
Y^2 = X^3 +S (T^3+aS^2T + bS^3)X^2 + S^7(cT+dS)X.
\end{equation}
It is a straightforward exercise to complete the cube and calculate the
discriminant, which is
\begin{equation}
\Delta = -S^{16}(cT+dS)^2 (T^6+2aS^2T^4+2bS^3T^3+a^2S^4T^2+(2ab-4c)S^5T+(b^2-4d)S^6).
\end{equation}
Since $S$ divides the coefficient of $X^2$ and $S^2$ divides the coefficient
of $X$ in \eqref{eq:homog}, we conclude that the fiber at $S=0$ is type $I_{10}^*$, so the
gauge algebra is enhanced to $\mathfrak{so}(28)$.  In addition, at
the point $[S,T]=[-c,d]$, the coefficient of $X^2$ is {\em not}\/ divisible
by $(cT+dS)$, so the Kodaira type is $I_2$ and there is an additional enhancement
of the gauge algebra to $\mathfrak{su}(2)$.  (For generic coefficients,
the other factor in the discriminant contributes six fibers of type $I_1$.)

Since the constant term in \eqref{eq:homog} vanishes, the section $X=Y=0$
defines an element of order  $2$ in the Mordell-Weil group.  It follows
as in \cite{MR1416960,pioneG} that the actual gauge group of this model
is $(\operatorname{Spin}(28)\times SU(2))/\mathbb{Z}_2$.

The intrinsic property of elliptically fibered K3 surfaces
which leads to an equation of the form \eqref{eq:so32} is the requirement
that there be a $2$-torsion element in the Mordell--Weil group, and that
one fiber in the fibration be of type $I_n^*$ for some $n\ge 10$.
Under these assumptions, we can choose coordinates so that the specified 
fiber is at $T=\infty$.  If we were to simply
ask that the fiber at $T=\infty$  be of type $I_{10}^*$ or worse (as
well as having a $2$-torsion element),
then a slight modification of the argument in section 4 of 
\cite{Aspinwall:1996vc} or appendix A or \cite{nongeometries}
 would show that the equation takes the form
\begin{equation} 
Y^2 = X^3 + (\alpha T^3 + \beta T^2 +aT + b)X^2 + (cT+d)X.
\end{equation}
However, if $\alpha=0$ then it turns out that the coefficient $f$ 
of the Weierstrass form vanishes to order
at least $3$ and the coefficient
$g$ vanishes to order at least $4$, which means that
the fiber no longer has type $I_n^*$.  Thus, our requirement of being
type $I_n^*$ for some $n\ge10$ implies that $\alpha\ne0$
(or in a family, that $\alpha$ has no zeros). Then the coordinate change
\begin{equation}
 (X,Y,T) \mapsto (\alpha^{-2} X, \alpha^{-3} Y, \alpha^{-1}(T-\frac13\beta))
\end{equation}
(followed by multiplying the equation by $\alpha^6$) yields an equation
of the form \eqref{eq:so32}, i.e., one in which
the coefficient of $T^3X^2$ is $1$ and the coefficient of $T^2X^2$ is $0$.

The lattice enhancements which we have discussed
can also be interpreted for these models.
When $c=0$, the gauge group enhances to $\operatorname{Spin}(32)/\mathbb{Z}_2$,
and when $q(a,b,c,d)=0$, there is an additional enhancement of the gauge
algebra by a factor of $\mathfrak{su}(2)$.

\section{Six-dimensional compactifications} \label{sec:sixD}

We now specialize to six-dimensional non-geometric heterotic compactifications.
The base $Z$ is a Riemann surface with an effective anti-canonical divisor,
so it must either be an elliptic curve or the Riemann sphere.  In the first
case $Z=T^2$, the line bundle $\Lambda$ is trivial, and the entire construction
is just a $T^2$ compactification of the eight-dimensional theory, with
no monodromy or brane insertions needed.  In particular,
the parameters of the eight-dimensional theory do not vary, and the
compactification is geometric.

More interesting is the case $Z=\mathbb{P}^1$.  In this case, as
derived at the end of Section~\ref{sec:nongeom}, we have 
$\Lambda=\mathcal{O}_{\mathbb{P}^1}(2)$,
$\Lambda^{\otimes 6}=\mathcal{O}_{\mathbb{P}^1}(12)$, and 
the base $W=\mathbb{P}(\mathcal{O}_{\mathbb{P}^1}
\oplus \mathcal{O}_{\mathbb{P}^1}(12))$ of the non-geometric model
coincides with the Hirzebruch surface 
$\mathbb{F}_{12}$, and is similar to models first studied in 
\cite{FCY2}. 
In particular, a Calabi-Yau three-fold $\mathbf{\bar{X}} \to \mathbb{F}_{12}$ can be defined by
the  Weierstrass equation 
\begin{equation}
\label{WE_MV2_2}
\begin{split}
 0 = - y^2 \, z +  x^3 +& \, s^4 \, t^3 \, \Big(a(u,v)  \, t +  c(u,v) \, s\Big) \, x \, z^2 \\
 + & \,  s^5 \, t^5 \, \Big( t^2 + b(u,v) \, s \,  t + d(u,v) \, s^2\Big) \, z^3 \;,
\end{split}
\end{equation}
where $[u:v]$ denotes the homogeneous coordinates for the $\mathbb{P}^1$ that constitutes the  base of the Hirzebruch surface $\mathbb{F}_{12}$, and $[s:t]$ denotes the homogeneous coordinates of the $\mathbb{P}^1$ that constitutes the fiber, and the coefficients $a(u,v)$, $b(u,v)$, $c(u,v)$, and $d(u,v)$ have degrees $8$, $12$, $20$, and $24$, respectively, as homogeneous polynomials on $\mathbb{P}^1$.
The two $\mathbb{C}^*$-torus actions
that define $\mathbb{F}_{12}$ are given by
\begin{equation}
 (s,t,u,v) \sim (\lambda^{-12} s, t, \lambda u , \lambda v) \;,\quad  (s,t,u,v) \sim (\lambda s, \lambda t, u , v) \;,
\end{equation} 
for $\lambda \in \mathbb{C}^*$. 
The model has a fiber of type $II^*$ over the section
$\sigma_\infty$ of self-intersection $-12$ given by $s=0$, and a fiber of type $III^*$
over a disjoint section $\sigma_0$ given by $t=0$ with $\sigma_0 = \sigma_\infty + 12\, F$ where $F$ is the fiber class.
The divisor class of $\Delta =0$ is $[\Delta]= - 12 K_{\mathbb{F}_{12}}$
where $K_{\mathbb{F}_{12}}= - 2 \, \sigma_\infty- 14 \, F$. The two curves $s=0$ and $t=0$ will account for a large portion 
of the divisor class $[\Delta]$. The remaining part $[\Delta']$ of the divisor not contained in $\sigma_0$ and $\sigma_\infty$
is $[\Delta'] = [\Delta] - 10 \, \sigma_\infty - 9 \, \sigma_0$.
It follows that what is left of the discriminant divisor $[\Delta']$ will not collide with $\sigma_\infty$ since
$\sigma_\infty \cdot [\Delta']=0$.  On the other hand,  the divisor $[\Delta']$ will collide with $\sigma_0$ in a total number of
$\sigma_0 \cdot [\Delta']=60$ points counted with multiplicity.  

Because of the choice of 
$\mathbb{F}_{12}$, from the $\mathfrak{e}_8\oplus \mathfrak{e}_7$
heterotic perspective
there are no instantons on the $\mathfrak{e}_8$ summand but the
$\mathfrak{e}_7$ summand has instanton number $24$.  Since instantons for
$\mathfrak{e}_8$ must
be pointlike instantons, this allows the heterotic model to make sense
perturbatively,\footnote{Note that this is a difference from
the case of unbroken gauge algebra $\mathfrak{e}_8 \oplus
\mathfrak{e}_8$ considered in \cite{nongeometries}, where all
instantons are pointlike no matter how the instanton numbers are
distributed between the two summands.} 
with instantons of finite size on $\mathfrak{e}_7$.

The enhancement from $\mathfrak{e}_7$ to $\mathfrak{e}_8$ thus occurs at the $20$ points
$\{c(z)=0\}$.  These points are regarded as responsible for the
matter representation of $\mathfrak{e}_7$
\cite{geom-gauge}, giving $20$ half-hypermultiplets in the
$56$-dimensional representation.

The enhancement from $\mathfrak{e}_8\oplus \mathfrak{e}_7$ to 
$\mathfrak{e}_8\oplus \mathfrak{e}_7\oplus \mathfrak{su}(2)$
occurs along the locus $q(a,b,c,d)=0$, which consists of $120$ points
on $\mathbb{P}^1$.  At these points, the $\mathbb{P}^1$ fiber of 
$\mathbb{F}_{12}$ is tangent to the residual discriminant divisor $\Delta'$.

Similarly, our requirement of a fiber of type $I_n^*$, $n\ge10$, in the 
$\mathfrak{so}(32)$ heterotic string will not allow for the
``hidden obstructor'' of \cite{Aspinwall:1996vc} to occur.
(Such ``hidden obstructor'' points occur when the coefficient of $T^3X^2$
in \eqref{eq:so32} vanishes; as previously discussed, this vanishing is
inconsistent with the fiber being of type $I_n^*$.)  
Avoiding these ``hidden obstructors'' allows
for a perturbative description in the $\mathfrak{so}(32)$ case as well.
The twenty zeros of $c(z)$ give rise to half-hypermultiplets in the
tensor product of the vector representation of $\operatorname{Spin}(28)$
with the fundamental representation of $\operatorname{SU}(2)$ (which is
again a $56$-dimensional quaternionic representation).

\section{Discussion}

Something rather surprising has just happened:  although we started with
a construction for non-geometric heterotic compactifications, the
resulting compactifications in six dimensions are actually the familiar
F-theory duals of geometric compactifications of the heterotic string
on K3 surfaces!  How did this happen?

Recall the original description of the heterotic/F-theory duality
in six dimensions, as described in \cite{FCY2} and
further amplified in \cite{Friedman:1997yq}:  the large radius limit
of the heterotic string corresponds to a degeneration limit of
the F-theory space, in which the F-theory base actually splits
into two components.  This limit involves tuning a holomorphic
parameter, and tuning holomorphic parameters away from constant
has an interesting property:  it is not possible to keep the parameter
controllably close to the limiting value.  Instead, once the holomorphically
varying quantity is non-constant, it samples {\em all}\/ values.

The conclusion, then, is that taking a heterotic compactification
even a ``small distance'' from the large radius limit destroys the
traditional semiclassical interpretation and no longer allows us to
discuss the compactification as being that of a manifold with a bundle.
This is not unlike what happens in type II compactifications, where
the analysis of $\Pi$-stability 
\cite{douglas,Douglas:2001hw,Aspinwall:2001dz}
shows that going any distance away from large radius limit, no matter how 
small, necessarily changes the stability conditions on some D-brane
classes and so destroys the semiclassical interpretation of the theory.

It would be interesting to check whether this phenomenon persists
in compactifications to four dimensions.  There, the presence of
fluxes may alter the structure of the moduli space, which is here
discussed using purely geometric considerations.  It may be that
when fluxes are involved, some truly new non-geometric models
can be constructed.  We leave this question for future work.

As pointed out to us by the referee, there may be interesting lessons
from this work for double field theory
(see, for example, 
\cite{Aldazabal:2013sca,Berman:2013eva,Hohm:2013bwa} and
references therein) and its heterotic extensions.  We leave this for
future work as well.

\subsection*{Acknowledgements}

We would like to thank Chuck Doran and Sav Sethi for helpful discussions.
The first author acknowledges the generous support of the
Kavli Institute for Theoretical Physics, and  the second author is grateful
to the Kavli Institute for the Physics and Mathematics of the Universe for
hospitality during the early stages of this project.
The work of the second author is supported by National Science Foundation
grants DMS-1007414 and PHY-1307513 and by the
World Premier International Research Center Initiative (WPI Initiative), MEXT, Japan.

\appendix

\section{Discrete groups and modular forms} \label{app:discrete}

\subsection{Modular forms for $O^+(\Lambda^{2,2})$}

Explicit generators for $O(\Lambda^{2,2})$ are given in \cite{MR2013800},
together with their actions on $\mathbb{H}\times \mathbb{H}$.
It implies that we can identify $O^+(\Lambda^{2,2})$ with
$P(\sltwoz \times \sltwoz)\rtimes \mathbb{Z}_2 $,
where the automorphism $\mathbb{Z}_2$ acts to exchange $\rho$ and $\tau$.

The modular forms of weight $d$
for this group must be functions of $\rho$ and $\tau$ of bidegree $(d,d)$
invariant under the exchange.  We claim that this ring of modular
forms is a free polynomial algebra on $E_4(\rho)E_4(\tau)$,
$E_6(\rho)E_6(\tau)$, and $\Delta_{12}(\rho)\Delta_{12}(\tau)$.

To see why this is true, let
 $t_4=E_4(\rho)E_4(\tau)$,
$t_6=E_6(\rho)E_6(\tau)$, and $t_{12}=\Delta_{12}(\rho)\Delta_{12}(\tau)$ 
be elements of bidegree $(4,4)$, $(6,6)$, and $(12,12)$, respectively.
If they are not algebraically independent, then there exists a nonvanishing homogeneous polynomial
$P(T)$ in the graded ring $\mathbb{C}[T_4,T_6,T_{12}]$ satisfying
$P(t_4,t_6,t_{12})=0$. We take as $P(T)$ the polynomial of minimal degree and write it
in the form $P_0(T_4,T_6,T_{12}) T_{12} + P_1(T_4,T_6)$.  In the equation
\[
 P_0(t_4,t_6,t_{12}) t_{12} + P_1(t_4,t_6) = 0 
\]
we take the limit $\rho \to i \infty$. Notice that under $\rho \to i \infty$ we have
\[
 E_4(\rho) \to 1 \;, \quad E_6(\rho) \to 1\; \quad \eta(\rho) \to 0 \;.
\]
Therefore, we get $P_1\big(E_4(\tau),E_6(\tau)\big)=0$. But $E_4(\tau)$ and $E_6(\tau)$ are algebraically independent. It follows that $P_1(T_4, T_6)=0$
and $P_0(T)$ is different from zero with $P_0(t_4,t_6,t_{12})=0$. Since $P_0(T)$ is
of smaller degree than $P(T)$ we get a contradiction to the  assumed minimality of $P(T)$.

Fix $k\ge 0$ even and let $n(k)$ denote the dimension of the space of modular forms $M_k$ of weight $k$ for
$\sltwoz$. It is well-known that 
\begin{equation}
\label{dimension}
n(k):= \dim M_k = \left\lbrace \begin{array}{ll} \lfloor k/12\rfloor & \text{for} \; k \equiv 2 (4), \\ \lfloor k/12 \rfloor+1 & \text{otherwise}. \end{array}\right. 
\end{equation}
Equivalently, the dimension $n(k)$ equals the number of nonnegative integer
solutions to the equation $k=4p+6q$ as $E_4$ and $E_6$ generate the ring of modular forms.  Then, the dimension of $M_k \otimes M_k$ is $n(k)^2$, and the dimension
of the linear subspace of bi-degree $(k,k)$ defined by $f(\rho,\tau)=f(\tau,\rho)$ is $\frac{1}{2}n(k)\big(n(k)+1\big)$. 
Let $R$ denote the graded subring generated by the algebraically independent $t_4, t_6, t_{12}$. The dimension of the subspace $R_k$ of bi-degree $(k,k)$
equals  the number of nonnegative integer solutions to the equation $k=4p+6q+12r$. If we fix $r$, then the number of such solutions
equals the dimension of $M_{k-12r}$. From Equation~(\ref{dimension}) it follows that for the dimension we have $\dim M_k = \dim M_{k-12} + 1$.
Let $r_0 =\lfloor k/12 \rfloor$ then summing over possible values for $r$ we obtain the dimension of $R_k$:
\[
 \dim R_k = \sum_{r=0}^{r_0} \dim M_{k-12r_0+12r} =  \left\lbrace \begin{array}{ll}    \sum_{r=1}^{r_0} r & \text{for} \, k \equiv 2 (4) \\[0.4em]  \sum_{r=0}^{r_0} r & \text{otherwise} \end{array} \right. \; = \frac{1}{2} \, n(k) \, \big(n(k)+1\big) \;,
 \]
 which agrees with the dimension of $\operatorname{Sym}^2(M_k)$.
 It follows that the elements $t_4$, $t_6$, $t_{12}$ generate $\operatorname{Sym}^2(M_*)$.

\subsection{The moduli spaces}
\label{app:modulispaces}

It is worthwhile to straighten out several moduli spaces of relevance
here. The key observation, due to Vinberg \cite{vinberg-siegel}, is that
under the natural homomorphism $\sptwor \to O^+(2,3)$, the
arithmetic group $\sptwoz$ (which is a maximal discrete subgroup
of $\sptwor$) maps to an index two subgroup $SO^+(L^{2,3}) \subset O^+(L^{2,3})$
where $O^+(2,3)$ denotes the subgroup of index 2 of $O(2,3)$ consisting of the elements whose spinor norm 
is equal to the determinant.
The isomorphism 
\begin{equation}
 \mathbb{H}_2/\sptwoz \cong \mathcal{D}_{2,3}/ SO^+(L^{2,3})
\end{equation}
gives rise to an isomorphism between the algebra of Siegel modular forms of genus 2 and 
the algebra of automorphic forms of 
$\mathcal{D}_{2,3}$
with respect to the group $SO^+(L^{2,3})$.
But the algebra of automorphic forms of 
$\mathcal{D}_{2,3}$ 
with respect to the group $O^+(L^{2,3})$
is the \emph{even} part of the algebra of automorphic forms with respect to $SO^+(L^{2,3})$ and, hence, is 
isomorphic to the algebra of \emph{even} Siegel modular forms of genus 2.

This means that the moduli space of principally polarized abelian surfaces
$\mathbb{H}_2/\sptwoz$ has a degree two map to the moduli space
of K3 surfaces with lattice polarization 
$\mathcal{D}_{2,3}/O^+(L^{2,3})$.
This can be understood at the level of modular forms as follows.  The Siegel
modular forms of even weight, generated by $\psi_4$, $\psi_6$, $\chi_{10}$
and $\chi_{12}$, are invariant under $O^+(L^{2,3})$ and give 
homogenous coordinates on the Baily-Borel compactification of that space.
On the other hand, the full ring of modular forms is invariant under
$\sptwoz$, and the equation $\chi_{35}^2=\mathcal{F}(\psi_4, \psi_6,\chi_{10},\chi_{12})$ 
(where $\mathcal{F}$ is given in Equation~(\ref{chi35sqr}))
expresses $\mathbb{H}_2/\sptwoz$
as a double cover of 
$\mathcal{D}_{2,3}/O^+(L^{2,3})$.

\begin{figure}
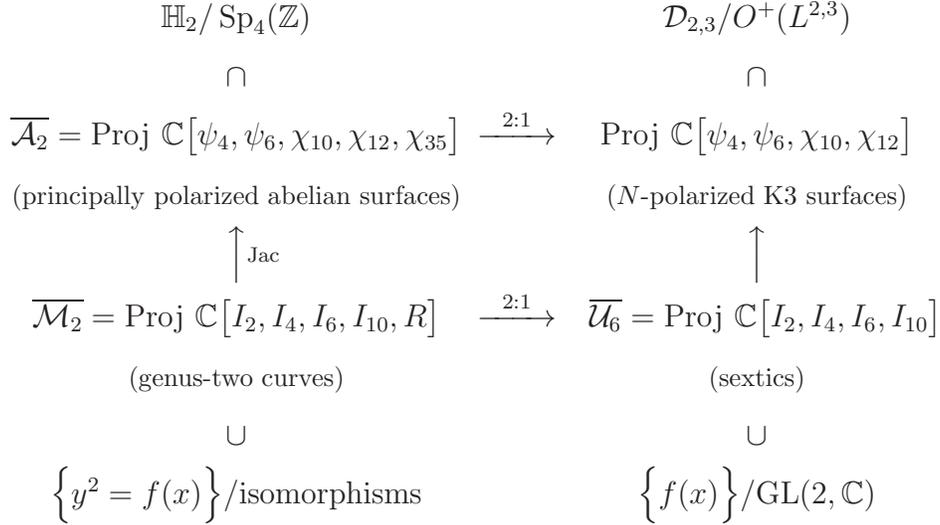

\begin{equation*}
\begin{CD}
\mathbb{H}_2/\sptwoz  @. 
\mathcal{D}_{2,3}/O^+(L^{2,3})
\\
\cap @. \cap\\
\overline{\mathcal{A}_2} = \operatorname{Proj} \, \mathbb{C}\big\lbrack \psi_4, \psi_6, \chi_{10}, \chi_{12}, \chi_{35} \big\rbrack @>2:1>>  \operatorname{Proj} \, \mathbb{C}\big\lbrack \psi_4, \psi_6, \chi_{10}, \chi_{12} \big\rbrack \\
\text{{\footnotesize (principally polarized abelian surfaces)}} @. \text{{\footnotesize ($N$-polarized K3 surfaces)}}\\
@AA\operatorname{Jac}A   @AAA\\
 \overline{\mathcal{M}_2} =  \operatorname{Proj} \, \mathbb{C}\big\lbrack I_2, I_4, I_6, I_{10}, R \big\rbrack @>2:1>>  \; \; \overline{\mathcal{U}_6}= \operatorname{Proj} \, \mathbb{C} \big\lbrack I_2, I_4, I_6, I_{10} \big\rbrack\\
\text{{\footnotesize (genus-two curves)}} @. \text{{\footnotesize (sextics)}}\\
\cup @. \cup\\
\Big \lbrace y^2 = f(x) \Big\rbrace / \text{isomorphisms} @.  \Big \lbrace f(x) \Big\rbrace / \mathrm{GL}(2,\mathbb{C})
\end{CD}
\end{equation*}
\caption{Relations among moduli spaces} \label{fig:moduli}
\end{figure}

This same phenomenon carries over to the moduli of genus two curves.
We can express a genus two curve in the form $y^2=f(x)$ (cf. Appendix~\ref{moduli_curves_genus2})
and so there is a map $\mathcal{M}_2 \to \mathcal{U}_6$ from the moduli space of
genus two curves to the moduli space of degree six polynomials or sextics.  This also
turns out to be a map of degree two.  The coordinates on the moduli space
of degree six polynomials were worked out by Clebsch: in Igusa's notation,
they are $I_2(f), I_4(f), I_6(f), I_{10}(f)$ and given in Equations~(\ref{IgusaClebschInvariants}).  
On the other hand, the moduli space of genus-two curves
has one additional invariant $R$ defined in Appendix \ref{moduli_curves_genus2}.
The point is that under the operation $f(x) \mapsto \tilde{f}(x)=f(-x)$ the odd invariant $R(f)$ is mapped to $R(\tilde{f})=-R(f)$
whereas the even invariants $I_2(f), I_4(f), I_6(f), I_{10}(f)$ remain the same, i.e.,  $I_{2k}(f)=I_{2k}(\tilde{f})$ for $k=2, 4,6 ,10$. 
The subtle point is that mapping $f \mapsto \tilde{f}$ and, hence, the ramification points $\theta_i \mapsto - \theta_i$ defines equivalent sextics,
but different genus-two curves. In fact, genus-two curves invariant under this action are the ones with bigger automorphism group with an extra involution of order two and $R(f)=0$.
Equation~(\ref{Rsqr}) expresses $R^2=\mathcal{F'}(I_2, I_4, I_6, I_{10})$ as a polynomial in terms of the \emph{even} Igusa-Clebsch invariants.
Therefore, this expresses $\mathcal{M}_2$ as a double cover of $\mathcal{U}_6$.
The  diagram in Figure~\ref{fig:moduli} summarizes the discussion.

\section{The moduli space of principally polarized abelian surfaces}
\label{app:B}

\subsection{The Siegel modular three-fold}
\label{SiegelThreefold}
The Siegel three-fold is a quasi-projective variety of dimension $3$ obtained from the Siegel upper 
half-plane of degree two which by definition is the set of two-by-two symmetric matrices over $\mathbb{C}$ whose imaginary part is positive definite, i.e.,
\begin{equation}
\label{Siegel_tau}
 \mathbb{H}_2 = \left. \left\lbrace \underline{\tau} = \left( \begin{array}{cc} \tau_1 & z \\ z & \tau_2\end{array} \right) \right|
 \tau_1, \tau_2, z \in \mathbb{C}\,,\; \im{(\tau_1)} \, \im{(\tau_2}) > \im{(z)}^2\,, \; \im{(\tau_2)} > 0 \right\rbrace \;,
\end{equation}
quotiented out by the action of the modular transformations $\Gamma_2:=
\sptwoz$, i.e., 
\begin{equation}
 \mathcal{A}_2 =  \mathbb{H}_2 / \Gamma_2 \;.
\end{equation} 
Each $\underline{\tau} \in \mathbb{H}_2$ determines a principally polarized complex abelian surface $\mathbf{A}_{\underline{\,\tau}} = \mathbb{C}^2 / \langle \mathbb{Z}^2 \oplus \underline{\tau} \,
\mathbb{Z}^2\rangle$ with period matrix $(\underline{\tau}, \mathbb{I}_2) \in \mathrm{Mat}(2, 4;\mathbb{C})$. 
Two abelian surfaces $\mathbf{A}_{\underline{\,\tau}}$ 
and $\mathbf{A}_{\underline{\,\tau}'}$ are isomorphic if and only if there is a symplectic matrix 
\begin{equation}
M= \left(\begin{array}{cc} A & B \\ C & D \end{array} \right) \in \Gamma_2
\end{equation}
such that $\underline{\tau}' = M (\underline{\tau}):=(A\underline{\tau}+B)(C\underline{\tau}+D)^{-1}$.
It follows that the Siegel three-fold $\mathcal{A}_2$ is also the set of isomorphism classes of principally polarized abelian surfaces.
The sets of abelian surfaces that have the same endomorphism ring form subvarieties of $\mathcal{A}_2$. 
The endomorphism ring of principally polarized abelian surface tensored with $\mathbb{Q}$ is either a quartic CM field, an indefinite
quaternion algebra, a real quadratic field or in the generic case $\mathbb{Q}$. Irreducible components of the corresponding 
subsets in $\mathcal{A}_2$ have dimensions $0, 1, 2$ and are known as CM points, Shimura curves and
Humbert surfaces, respectively. 

The Humbert surface $H_{\Delta}$ with invariant $\Delta$ is the space of principally polarized abelian surfaces admitting a symmetric endomorphism 
with discriminant $\Delta$. It turns out that $\Delta$ is a positive integer $\equiv 0, 1\mod 4$. In fact, $H_{\Delta}$ is the image inside $\mathcal{A}_2$ 
under the projection of the rational divisor associated to the equation
\begin{equation}
 a \, \tau_1 + b \, z + c \, \tau_3 + d\, (z^2 -\tau_1 \, \tau_2) + e = 0 \;,
\end{equation}
with integers $a, b, c, d, e$ satisfying $\Delta=b^2-4\,a\,c-4\,d\,e$ and $\underline{\tau}
= \bigl(\begin{smallmatrix}
\tau_1&z\\ z&\tau_2
\end{smallmatrix} \bigr) \in \mathbb{H}_2$.
For example, inside of $\mathcal{A}_2$ sit the Humbert surfaces $H_1$ and $H_4$ that are defined as the images under the projection 
of the rational divisor associated to $z=0$ and $\tau_1=\tau_2$, respectively. Equivalently, these points are invariant under the $\mathbb{Z}_2$-action generated by 
$ \bigl(\begin{smallmatrix}
A&0\\ 0&A
\end{smallmatrix} \bigr)  \in \Gamma_2$ with $A=\bigl(\begin{smallmatrix}
0&1\\ 1&0
\end{smallmatrix} \bigr)$ and  $A=\bigl(\begin{smallmatrix}
1&0\\ 0&-1
\end{smallmatrix} \bigr)$, respectively.
In fact, the singular locus of $\mathcal{A}_2$ has $H_1$ and $H_4$ as its two connected components. 
As analytic spaces, the surfaces $H_1$ and $H_4$ are each isomorphic to the Hilbert modular surface 
\begin{equation}
\label{modular_product2}
 \Big( (\sltwoz \times \sltwoz) \rtimes \mathbb{Z}_2 \Big) \backslash \Big( \mathbb{H} \times \mathbb{H} \Big) \;.
\end{equation}
For a more detailed introduction to Siegel modular form, Humbert surfaces, and the Satake compactification of the Siegel modular threefold we refer to
Freitag's book \cite{MR871067}.

\subsection{Siegel modular forms}
\label{Siegel_modular_forms}
In general, we can define the Eisenstein series $\psi_{2k}$ of degree $g$ and weight $2k$ (where we assume $2k>g+1$ for convergence)
by setting
\begin{equation}
 \psi_{2k}(\underline{\tau}) = \sum_{(C,D)} \det(C\cdot\underline{\tau}+D)^{-2k} \;,
\end{equation}
where the sum runs over non-associated bottom rows $(C,D)$ of elements in 
$\spgz$ where non-associated means with respect to the multiplication on the
left by $\mathrm{GL}(g,\Z)$. For $g=1$ and $k>1$, we have 
$\sponez=
\sltwoz$, $\operatorname{GL}(1,\Z)=\Z_2$, and
we obtain $\psi_{2k}(\tau)=E_{2k}(\tau)$ where $E_{2k}(\tau) = 1 + O(q)$ with $q=\exp{(2\pi i \tau)}$ are the standard normalized 
Eisenstein series.
The reason is that the series $E_{2k}$ be written as  
\begin{equation}
 E_{2k}(\tau) = \sum_{\substack{(c,d) = 1\\ (c,d)\equiv (-c,-d)} } \frac{1}{(c \tau+d)^{2k}} \;,
\end{equation}
where  the sum runs over all pairs of co-prime integers up to simultaneous $\Z_2=\mathrm{GL}(1,\Z)$ action.
The connection to the Eisenstein series $G_{2k}$ is given by
\begin{equation}
 G_{2k}(\tau) = 2 \, \zeta(2k) \, E_{2k}(\tau) 
 = \sum_{(m,n)\in \mathbb{Z}^2\backslash(0,0)} \frac{1}{(m\tau+n)^{2k}} \;. 
\end{equation}
In the following, we will always assume $g=2$ in the definition of $\psi_{2k}$.
For $z\to 0$, we then have that
\begin{equation}
\label{relation1}
 \psi_{2k} \left( \begin{array}{cc} \tau_1 & z \\ z & \tau_2 \end{array}\right) = E_{2k}(\tau_1) \; E_{2k}(\tau_2) + O(z^2)\;.
\end{equation}
Following Igusa \cite{MR0141643} we define a cusp form of weight $12$ by
\begin{equation}
\label{definition2}
\begin{split}
 \chi_{12}(\underline{\tau}) & = \frac{691}{2^{13}\, 3^8\, 5^3\, 7^2}\left(3^2\, 7^2 \, \psi_4^3(\underline{\tau})+2\cdot 5^3 \, \psi_6^2(\underline{\tau})-691\, \psi_{12}(\underline{\tau})\right) \;.
\end{split}
\end{equation}
We find that for $z\to 0$ its asymptotic behavior is given by
\begin{equation}
\label{relation2}
\begin{split}
 \chi_{12} \left( \begin{array}{cc} \tau_1 & z \\ z & \tau_2 \end{array}\right) & = \eta^{24}(\tau_1) \; \eta^{24}(\tau_2) + O(z^2) 
\end{split}
\end{equation}
where $\eta(\tau)$ is the Dedekind $\eta$-function and we have used that
\begin{equation}
\begin{split}
 1728 \, \eta^{24}(\tau_j) &= E_4^3(\tau_j) - E_6^2(\tau_j) \;,\\
 691 \, E_{12}(\tau_j) & = 441 \, E_{4}(\tau_j)^3 + 250 \, E_{6}(\tau_j)^2 \;.
\end{split}
\end{equation}
Igusa's `original' definition \cite[Sec.~8, p.~195]{MR0141643} for $\chi_{12}$ is 
\begin{equation}
\label{definition2b}
\begin{split}
 \tilde{\chi}_{12}(\underline{\tau}) & = \frac{131\cdot 593}{2^{13}\, 3^7\, 5^3\, 7^2 \, 337}\left(3^2\, 7^2 \, \psi_4^3(\underline{\tau})+2\cdot 5^3 \, \psi_6^2(\underline{\tau})-691\, \psi_{12}(\underline{\tau})\right) \\
 & = \frac{3 \cdot 131 \cdot 593}{337 \cdot 691} \, \chi_{12}(\underline{\tau}) = 1.00078\dots \; \chi_{12}(\underline{\tau})\;.
\end{split}
\end{equation}
But all results in \cite{MR0141643} that connect $\chi_{12}$ to the Igusa-Clebsch coefficients are based on the asymptotic expansion in Equation (\ref{relation2}). Hence, the definition
in Equation~(\ref{definition2}) must be used. Using Igusa's definition \cite[Sec.~8, p.~195]{MR0141643} we also define a second cusp form of weight $10$ by
\begin{equation}
\begin{split}
 \chi_{10}(\underline{\tau}) & = - \frac{43867}{2^{12}\, 3^5\, 5^2\, 7 \cdot 53} \left(\psi_4(\underline{\tau}) \, \psi_6(\underline{\tau}) - \psi_{10}(\underline{\tau})\right) \;.
\end{split}
\end{equation}
We see that for $z\to 0$ its asymptotic behavior is given by
\begin{equation}
\label{relation3}
\begin{split}
 \chi_{10} \left( \begin{array}{cc} \tau_1 & z \\ z & \tau_2 \end{array}\right) & = \eta^{24}(\tau_1) \; \eta^{24}(\tau_2) \, (\pi\,z)^2 + O(z^4) \;.
\end{split}
\end{equation}
Hence, the vanishing divisor of the cusp form $\chi_{10}$ is the Humbert surface $H_1$ because
a period point $\underline{\tau}$ is equivalent to a point with $z=0$ if and only if $\chi_{10}\big(\underline{\tau}\big)=0$.

Igusa proved \cite{MR0229643, MR527830} that the ring of Siegel modular forms is generated by
$\psi_4$, $\psi_6$, $\chi_{10}$, $\chi_{12}$ and by one more cusp form $\chi_{35}$ of odd weight $35$ 
whose square is the following polynomial \cite[p.~849]{MR0229643} in the even generators 
\begin{equation}\label{chi35sqr}
\begin{split}
\chi_{35}^2 & = \frac{1}{2^{12} \, 3^9} \; \chi_{10} \,  \Big(  
2^{24} \, 3^{15} \; \chi_{12}^5 - 2^{13} \, 3^9 \; \psi_4^3 \, \chi_{12}^4 - 2^{13} \, 3^9\; \psi_6^2 \, \chi_{12}^4 + 3^3 \; \psi_4^6 \, \chi_{12}^3 \\
& - 2\cdot 3^3 \; \psi_4^3 \, \psi_6^2 \, \chi_{12}^3 - 2^{14}\, 3^8 \; \psi_4^2 \, \psi_6 \, \chi_{10} \, \chi_{12}^3 -2^{23}\, 3^{12} \, 5^2\, \psi_4 \, \chi_{10}^2 \, \chi_{12}^3  + 3^3 \, \psi_6^4 \, \chi_{12}^3\\
& + 2^{11}\,3^6\,37\,\psi_4^4\,\chi_{10}^2\,\chi_{12}^2+2^{11}\,3^6\,5\cdot 7 \, \psi_4 \, \psi_6^2\, \chi_{10}^2 \, \chi_{12}^2 -2^{23}\, 3^9 \, 5^3 \, \psi_6\, \chi_{10}^3 \, \chi_{12}^2 \\
& - 3^2 \, \psi_4^7 \, \chi_{10}^2 \, \chi_{12} + 2 \cdot 3^2 \, \psi_4^4 \, \psi_6^2 \, \chi_{10}^2 \, \chi_{12} + 2^{11} \, 3^5 \, 5 \cdot 19 \, \psi_4^3 \, \psi_6 \, \chi_{10}^3 \, \chi_{12} \\
&  + 2^{20} \, 3^8 \, 5^3 \, 11 \, \psi_4^2 \, \chi_{10}^4 \, \chi_{12} - 3^2 \, \psi_4 \, \psi_6^4 \, \chi_{10}^2 \, \chi_{12} + 2^{11} \, 3^5 \, 5^2 \, \psi_6^3 \, \chi_{10}^3 \, \chi_{12}  - 2 \, \psi_4^6 \, \psi_6 \, \chi_{10}^3 \\
 & - 2^{12} \, 3^4 \, \psi_4^5 \, \chi_{10}^4 + 2^2 \, \psi_4^3 \, \psi_6^3 \, \chi_{10}^3 + 2^{12} \, 3^4 \, 5^2 \, \psi_4^2 \, \psi_6^2 \, \chi_{10}^4 + 2^{21} \, 3^7 \, 5^4 \, \psi_4 \, \psi_6 \, \chi_{10}^5 \\
 & - 2 \, \psi_6^5 \, \chi_{10}^3 + 2^{32} \, 3^9 \, 5^5 \, \chi_{10}^6 \Big) \;.
\end{split}
\end{equation}
Hence, $Q:= 2^{12} \, 3^9 \, \chi_{35}^2 /\chi_{10}$ is a polynomial of degree $60$
in the even generators. One then checks that
\begin{equation}
 q\left(-\frac1{48}\psi_4,-\frac1{864}\psi_6,-4 \, \chi_{10}, \, \chi_{12}\right)  =  \frac{1}{2^{20} \, 3^9} \, Q(\psi_4, \psi_6, \chi_{10}, \chi_{12})  \;,
\end{equation}
where $q$ was defined in Equation (\ref{q-eqn}). 
It is known that the vanishing divisor of $Q$ is the 
Humbert surface $H_4$ \cite{MR1438983} because
a period point $\underline{\tau}$ is equivalent to a point with $\tau_1=\tau_2$ if and only if $Q\big(\underline{\tau}\big)=0$.
Accordingly, the vanishing divisor of $\chi_{35}$ is the formal sum $H_1 + H_4$ of Humbert surfaces, that constitutes the singular locus of $\mathcal{A}_2$.

In accordance with Igusa \cite[Theorem 3]{MR0141643} we also introduce the following ratios of Siegel modular forms
\begin{equation}
\label{ratio_of_Siegel_forms}
 \mathbf{x}_1 = \dfrac{\psi_4 \, \chi_{10}^2}{\chi_{12}^2} ,\quad
 \mathbf{x}_2 = \dfrac{\psi_6\, \chi_{10}^3}{\chi_{12}^3} ,\quad
 \mathbf{x}_3 = \dfrac{\chi_{10}^6}{\chi_{12}^5} \;,
\end{equation}
as well as
\begin{equation}
\label{ratio_of_Siegel_forms2}
 \mathbf{y}_1 = \dfrac{\mathbf{x}_1^3}{\mathbf{x}_3} = \dfrac{\psi_4^3}{\chi_{12}} ,\quad
 \mathbf{y}_2 = \dfrac{\mathbf{x}_2^2}{\mathbf{x}_3} = \dfrac{\psi_6^2}{\chi_{12}} ,\quad
 \mathbf{y}_3 = \dfrac{\mathbf{x}_1^2 \, \mathbf{x}_2}{\mathbf{x}_3} = \dfrac{\psi_4^2 \, \psi_6 \, \chi_{10}}{\chi_{12}} \;,
\end{equation}
where we have suppressed the dependence of each Siegel modular form on $\underline{\tau}$.
These ratios have the following asymptotic expansion as $z\to 0$ \cite[pp.~180--182]{MR0141643}
\begin{equation}
\label{asymptotics2}
\begin{split}
 \mathbf{x}_1 & = E_4(\tau_1) \; E_4(\tau_2) \, (\pi z)^{4} + O(z^{5}) \;, \\
 \mathbf{x}_2 & = E_6(\tau_1) \; E_6(\tau_2) \, (\pi z)^{6} + O(z^{7}) \;, \\
 \mathbf{x}_3 & =  \eta^{24}(\tau_1) \; \eta^{24}(\tau_2) \, (\pi z)^{12} + O(z^{13})\;,\\
\end{split}
\end{equation}
and
\begin{equation}
\label{asymptotics3}
\begin{split}
 \mathbf{y}_1 & = j(\tau_1) \, j(\tau_2) + O(z^2) \;, \\
 \mathbf{y}_2 & = \Big(1728 - j(\tau_1)\Big) \, \Big(1728 -j(\tau_2)\Big) + O(z^2) \;, \\
 \mathbf{y}_3 & = \dfrac{E_4^2(\tau_1) \, E_4^2(\tau_2) \, E_6(\tau_1) \, E_6(\tau_2)}{\eta^{24}(\tau_1) \, \eta^{24}(\tau_2)} \, (\pi z)^2 + O(z^3)\;,
\end{split}
\end{equation}
where we have set
\begin{equation}
\label{j_invariant}
\begin{split}
 j(\tau_j) & = \dfrac{1728 \, E_4^3(\tau_j)}{E_4^3(\tau_j)-E_6^2(\tau_j)} =\dfrac{E_4^3(\tau_j)}{\eta^{24}(\tau_j)}\;,  \\
 1728 - j(\tau_j) & = \dfrac{1728 \, E_6^2(\tau_j)}{E_4^3(\tau_j)-E_6^2(\tau_j)} =\dfrac{E_6^2(\tau_j)}{\eta^{24}(\tau_j)}\;.
 \end{split}
\end{equation}
Notice that for the asymptotic behavior in Equations (\ref{asymptotics2}) and (\ref{asymptotics3}) the right normalization of $\chi_{12}$ was essential.

\subsection{Sextics and Igusa invariants}
\label{moduli_curves_genus2}
We write the equation defining a genus-two curve $C$ by a degree-six polynomial or sextic in the form
\begin{equation}
\label{genus_two_curve}
 C: \; y^2 = f(x)  = a_0 \, \prod_{i=1}^6 (x-\theta_i) 
 = \sum_{i=0}^6 a_i \, x^{6-i}\;.
\end{equation}
The roots $(\theta_i)_{i=1}^6$ of the sextic are the six ramification points of the map $C \to \mathbb{P}^1$. Their
pre-images on $C$ are the six Weierstrass points. The isomorphism class of $f$ consists of all equivalent
sextics where two sextics are considered equivalent if there is a linear transformation in $\mathrm{GL}(2,\mathbb{C})$ which takes the
set of roots to the roots of the other. The action of the linear transformations on the Weierstrass points defines a 7-dimensional irreducible linear representation 
of $\sltwoc$. The corresponding invariants are called the invariants of the sextic. 

Clebsch defined such invariants $I_2, I_4, I_6, I_{10}$ of weights $2, 4, 6, 10$, respectively,  now
called the \emph{Igusa-Clebsch invariants} of the sextic curve in (\ref{genus_two_curve}), as follows
\begin{equation}
\label{IgusaClebschInvariants}
\begin{split}
 I_2(f) & = a_0^2 \, \sum_{i<j, k<l, m<n} D^2_{ij} \, D^2_{kl} \, D^2_{mn} \;, \\
 I_4(f) & = a_0^4 \sum_{\substack{i<j<k, l < m <n}} D^2_{ij} \, D^2_{jk} \, D^2_{ki} \, D^2_{lm} \, D^2_{mn} \, D^2_{nl}\;,\\
 I_6(f) & = a_0^6 \sum_{\substack{i<j<k, l<m<n\\i<l', j<m', k<n'\\l', m', n' \in \lbrace l, m, n \rbrace}} D^2_{ij} \, D^2_{jk} \, D^2_{ki} \, D^2_{lm} \, D^2_{mn} \, D^2_{nl} \, D^2_{il'} \, D^2_{jm'} \, D^2_{kn'} \;,\\
 I_{10}(f) & = a_0^{10} \, \prod_{i<j} D^2_{ij}  \;,
\end{split}
\end{equation} 
where  $D_{ij} =\theta_i - \theta_j$ and all indices take values in $\lbrace 1, \dots, 6\rbrace$. In the following, we will often suppress the argument $f$.
The invariants $(I_2 , I_4 , I_6 , I_{10})$ are the same invariants as $( A' , B' , C' , D')$ in \cite[p.~319]{MR1106431} and also the same invariants as $(A, B, C, D)$ in \cite[p.~176]{MR0141643}. It follows from the work of Mestre \cite{MR1106431} that the Igusa-Clebsch invariants
are arithmetic invariants, i.e., polynomials in the coefficients $a_0, \dots, a_6$ with integer coefficients, or $I_k \in \mathbb{Z}[a_0,\dots,a_6]$ for $k\in \{2, 4, 6, 10\}$.
Furthermore, a theorem by Bolza and Clebsch states that two sextics given by $f$ and $f'$ are isomorphic if and only if 
there is a $\rho \in \mathbb{C}^*$ such that $I_{2k}(f')=\rho^{-2k}\, I_{2k}(f)$ for $k=1,2,3,5$. Thus, the invariants of a sextic
define a point in a weighted projective space
$[I_2 : I_4 : I_6 : I_{10}] \in \mathbb{WP}^3_{(2,4,6,10)}$. It was shown in \cite{MR0141643} that points in the projective 
variety $\operatorname{Proj}\, \mathbb{C} [I_2, I_4, I_6, I_{10}]$ which
are not on $I_{10}=0$ form the variety $\mathcal{U}_6 $ of moduli of sextics.
Equivalently, points in this weighted projective space $\{[I_2 : I_4 : I_6 : I_{10}] \in \mathbb{WP}^3_{(2,4,6,10)}:
I_{10} \not = 0\}$ are in one-to-one correspondence with isomorphism classes of sextics.  

Often the \emph{Clebsch invariants} of a sextic are used as well. The Clebsch invariants $(A,B,C,D)$ are
related to the Igusa-Clebsch invariants by the equations
\begin{equation}\label{Clebsch_invariants}
\begin{split}
 I_2  = &\; -120 \, A\;, \\
 I_4  = & \; -720 \, A^2 + 6750 \, B\;,\\
 I_6  = & \; \phantom{X;} 8640 \, A^3 - 108000 \, A \, B + 202500 \, C\;,\\
 I_{10} = & \; -62208 \, A^5 + 972000 \, A^3 \, B + 1620000 \, A^2 \, C \\
 & \; - 3037500 \, A \, B^2 - 6075000 \, B \, C - 4556250 \, D \;.
\end{split}
\end{equation} 
Conversely, the invariants $(A,B,C,D)$ are polynomial expressions in the Igusa invariants $(I_2, I_4, I_6, I_{10})$ with rational coefficients.
Mestre \cite{MR1106431} also defined the following polynomials in the Clebsch invariants
\begin{equation}\label{Clebsch_invariants2}
\begin{split}
 A_{11} & = 2 \, C + \frac{1}{3} \, A\, B\;, \\
 A_{22} & = A_{31} = \, D\;,\\
 A_{33} & = \frac{1}{2} \, B \, D + \frac{2}{9} \, C \, (B^2+ A \, C) \;,\\
 A_{23} & = \frac{1}{3} \, B \, (B^2+ A \, C) + \frac{1}{3} \, C \, (2\, C + \frac{1}{3} \, A \, B) \;,\\
 A_{12} & = \frac{2}{3} \, (B^2+ A\, C) \;.
\end{split}
\end{equation} 
According to \cite{MR1106431} one can obtain from a sextic $f$ three binary quadrics of the form
\begin{equation}
  \mathsf{y}_i(x) := \alpha_i \, x^2 + \beta_i \, x + \gamma_i  
\end{equation}
with $i=1,2,3$ by an operation called `\"Uberschiebung' \cite[p.~317]{MR1106431}. To fix the normalization and order of the quadrics we remark that 
in the notation of \cite{MR1106431} we have $I_{10}=(\mathsf{y}_3\mathsf{y}_1)_2$.
The quadrics  $\mathsf{y}_i$ for $i=1,2,3$ have the property that their coefficients are polynomial expressions in the coefficients of $f$ with rational coefficients.
Moreover, under the operation $f(x) \mapsto \tilde{f}(x)=f(-x)$
the quadrics change according to $\mathsf{y}_i(x) \mapsto \tilde{\mathsf{y}}_i(x)= \mathsf{y}_i(-x)$ for $i=1,2,3$. Hence,
they are not invariants of the sextic. In contrast, $I_2(f), I_4(f), I_6(f), I_{10}(f)$ remain unchanged under this operation, 
i.e.,  $I_{2k}(f)=I_{2k}(\tilde{f})$ for $k=2, 4,6 ,10$. This latter statement is
easily checked since Equations~(\ref{IgusaClebschInvariants}) are invariant under $f \mapsto \tilde{f}$ or, equivalently, $D_{ij} \mapsto - D_{ij}$.

We define $R$ to be $1/4$ times the determinant of the three binary quadrics $ \mathsf{y}_i$ for $i=1,2,3$ 
with respect to the basis $x^2, x, 1$. It is obvious that under the operation $f(x) \mapsto \tilde{f}(x)=f(-x)$
the determinant $R$ changes its sign, i.e.,  $R(f) \mapsto R(\tilde{f})=-R(f)$.
A calculation shows that
\begin{equation}\label{Rsqr}
 R^2 = \frac{1}{2} \, \left| \begin{array}{ccc} A_{11} & A_{12} & A_{31} \\ A_{12} & A_{22} & A_{23} \\ A_{31} & A_{23} & A_{33} \end{array}\right| \;,
\end{equation} 
where $A_{ij}$ are the Clebsch invariants~(\ref{Clebsch_invariants2}). Like $R(f)^2$, the coefficients $A_{ij}(f)$ are invariant under the operation 
$f(x) \mapsto \tilde{f}(x)=f(-x)$ as they are polynomials in $(I_2, I_4, I_6, I_{10})$. 
Bolza \cite{MR1505464} described the possible automorphism groups of genus-two curves defined by sextics. In particular, he provided effective criteria for 
the cases when the automorphism group of the sextic curve in (\ref{genus_two_curve}) is nontrivial. The results are as follows:
\begin{enumerate}
\item The curve has an extra involution other than the exchange of sheets $(x,y) \to (x,-y)$ if and only if $R^2=0$. 
The sextic is then isomorphic to $f(x)=x^6 + c_1 \, x^4 + c_2 \, x^2 + 1$ for some $c_1, c_2 \in \mathbb{C}$ with the extra involution $(x,y) \to (-x,y)$.
\item The automorphism group contains an element of order $5$ if and only if $I_2=I_4=I_6=0,I_{10}\not = 0$. The sextic is then isomorphic to $f(x)=x(x^5+1)$ with 
the element of order $5$ being $(x,y) \to (\zeta_5^{2} \, x, \zeta_5 y)$ where $\zeta_5 = \exp{(2\pi i/5)}$.
\end{enumerate}

\subsection{The moduli space of genus-two curves}
\label{moduli_space_chi12}
Suppose that $C$ is an irreducible projective nonsingular curve. If the self-intersection is $C\cdot C=2$ then $C$ is a curve of genus two.
For every curve $C$ of genus two there exists a unique pair $(\mathrm{Jac}(C),j_C)$ where $\mathrm{Jac}(C)$ is an abelian surface, called the
Jacobian variety of the curve $C$, and $j_C: C \to \mathrm{Jac}(C)$ is an embedding. One can always regain $C$ from the pair $(\mathrm{Jac}(C),\mathcal{P})$ where
$\mathcal{P}=[C]$ is the class of $C$ in the N\'eron-Severi group $\mathrm{NS}(\mathrm{Jac}(C))$. 
Thus, if $C$ is a genus-two curve, then $\mathrm{Jac}(C)$ is a principally polarized abelian surface
with principal polarization $\mathcal{P}=[C]$, and the map sending a curve $C$ to its Jacobian variety $\mathrm{Jac}(C)$ is injective. 
In this way, the variety of moduli of curves of genus two is also the moduli space of their Jacobian varieties with canonical polarization. 
Since we have $\mathcal{P}^2=2$, the transcendental lattice is $\mathrm{T}(\mathrm{Jac}(C)) = \Lambda^{2,2} \oplus \langle -2 \rangle$.
Furthermore, Torelli's theorem states that the map sending a curve $C$ to its Jacobian variety $\mathrm{Jac}(C)$ induces a birational map from the moduli space $\mathcal{M}_2$ 
of genus-two curves to the complement of the Humbert surface $H_1$ in $\mathcal{A}_2$, i.e., $\mathcal{A}_2 - \supp{(\chi_{10})}_0$.

One can then ask what the Igusa-Clebsch invariants of a genus-two curve $C$ defined by a sextic curve $f$ are in terms of $\underline{\tau}$ 
such that $(\underline{\tau}, \mathbb{I}_2) \in \mathrm{Mat}(2, 4;\mathbb{C})$ is the period matrix of the principally polarized abelian surface 
$\mathbf{A}_{\underline{\tau}}=\mathrm{Jac}(C)$. Based on the asymptotic behavior in Equations (\ref{asymptotics2}) and (\ref{asymptotics3}), Igusa \cite{MR0141643} proved that the relations are as follows:
\begin{equation}
\label{invariants}
\begin{split}
 I_2(f) & = \dfrac{\chi_{12}(\underline{\tau})}{\chi_{10}(\underline{\tau})} \;, \\
 I_4(f) & = \frac{1}{2^4 \, 3^2} \, \psi_4(\underline{\tau}) \;,\\
 I_6(f) & = \frac{1}{2^6 \, 3^4} \, \psi_6(\underline{\tau}) + \frac{1}{2^4 \, 3^3} \,  \dfrac{\psi_4(\underline{\tau}) \, \chi_{12}(\underline{\tau})}{\chi_{10}(\underline{\tau})} \;,\\
 I_{10}(f) & = \frac{1}{2 \cdot 3^5} \, \chi_{10}(\underline{\tau}) \;.
\end{split}
\end{equation}
Thus, we find that the point $[I_2 : I_4 : I_6 : I_{10}]$ in weighted projective space equals
\begin{equation}
\label{IgusaClebschProjective}
\begin{split}
 \Big[ 2^3 \, 3 \, (3r\chi_{12})\, : \, 2^2 3^2 \, \psi_4 \,  (r\chi_{10})^2 \, : \, 2^3\, 3^2\, \Big(4 \psi_4  \, (3r\chi_{12})+ \psi_6 \,(r\chi_{10}) \Big)\, (r\chi_{10})^2: 2^2 \,  (r\chi_{10})^6 \Big]
 \end{split}
\end{equation}
with $r=2^{12}\, 3^5$. 
Substituting (\ref{invariants}) into Equations (\ref{Clebsch_invariants}), (\ref{Clebsch_invariants2}) it also follows that
\begin{equation}
\begin{split}
   R(f)^2 & = 2^{-41} \, 3^{-42} \, 5^{-20} \;
   \dfrac{Q\Big(\psi_4(\underline{\tau}), \psi_6(\underline{\tau}), \chi_{10}(\underline{\tau}), \chi_{12}(\underline{\tau})\Big)}{\chi_{10}(\underline{\tau})^3} \\    
   & =   2^{-29} \, 3^{-33} \, 5^{-20}\; \dfrac{\chi_{35}(\underline{\tau})^2}{\chi_{10}(\underline{\tau})^4}\;,
\end{split}
\end{equation}
where $Q$ and $R^2$ where defined in Equation (\ref{chi35sqr}) and (\ref{Rsqr}), respectively.

If $\underline{\tau}$ is equivalent to a point with $\tau_1=\tau_2$ or $[\underline{\tau}] \in H_4 \subset \mathcal{A}_2$
then the corresponding sextic curve has an extra automorphism with $R(f)^2=0$.
The transcendental lattice degenerates to $\mathrm{T}(\mathbf{A}_{\underline{\tau}}) = \Lambda^{1,1} \oplus \langle 2 \rangle \oplus \langle -2 \rangle$. 
If $\underline{\tau}$ is equivalent to a point with $z=0$ or $[\underline{\tau}] \in H_1 \subset \mathcal{A}_2$,
then the principally polarized abelian surface is a product of two elliptic curves  $\mathbf{A}_{\underline{\tau}}=E_{\tau_1} \times E_{\tau_2}$
because of Equations (\ref{weighted_projective_coords}) and (\ref{asymptotics2}). 
The transcendental lattice degenerates to $\mathrm{T}(\mathbf{A}_{\underline{\tau}}) = \Lambda^{2,2}$. 

For $I_2 \not =0$ we use the variables $\mathbf{x}_1, \mathbf{x}_2, \mathbf{x}_3$ from Equations~(\ref{ratio_of_Siegel_forms}) to write
\begin{equation}
\label{weighted_projective_coords}
\begin{split}
 \Big\lbrack I_2 : I_4 : I_6 : I_{10} \Big\rbrack =   \left\lbrack 1 : \frac{1}{2^4 \,3^2} \,  \mathbf{x}_1 : \frac{1}{2^6 \, 3^4} \, \mathbf{x}_2 + \frac{1}{2^4 \, 3^3} \, \mathbf{x}_1
 :  \frac{1}{2 \cdot 3^5} \, \mathbf{x}_3\right\rbrack \in \mathbb{WP}^3_{(2,4,6,10)} \;.
\end{split}
\end{equation}
Since the invariants $I_4, I_6, I_{10}$ vanish simultaneously at sextics with triple roots all such abelian surfaces are mapped to $[1:0:0:0] \in \mathbb{WP}^3_{(2,4,6,10)}$ with uniformizing affine coordinates $\mathbf{x}_1, \mathbf{x}_2, \mathbf{x}_3$ around it. Blowing up this point gives a variety that 
parameterizes genus-two curves with $I_2 \not = 0$ and their degenerations.  
In the blow-up space we have to introduce additional coordinates that are obtained as ratios  of $\mathbf{x}_1, \mathbf{x}_2, \mathbf{x}_3$ and have weight zero. Those are precisely the coordinates $\mathbf{y}_1, \mathbf{y}_2, \mathbf{y}_3$ already introduced in Equation~(\ref{ratio_of_Siegel_forms2}). It turns out that the coordinate ring of the blown-up space is $\mathbb{C}[\mathbf{x}_1, \mathbf{x}_2, \mathbf{x}_3, \mathbf{y}_1, \mathbf{y}_2, \mathbf{y}_3]$. 

If a Jacobian variety corresponds to a product of elliptic curves then $\underline{\tau}$ is equivalent to a point with $z=0$, i.e., 
$\underline{\tau}$ is located on the Humbert surface $H_1$. We then have
$\chi_{10}(\underline{\tau})=0, \chi_{12}(\underline{\tau})\not = 0$ and $[I_2:I_4:I_6:I_{10}]=[1:0:0:0]$.
Equations~(\ref{asymptotics2}) and  (\ref{asymptotics3}) imply $\mathbf{x}_1 = \mathbf{x}_2 = \mathbf{x}_3 = \mathbf{y}_3 = 0$ 
and  $\mathbf{y}_1  = j(\tau_1) \, j(\tau_2) $ and $ \mathbf{y}_2  = (1728 - j(\tau_1)) \, (1728 -j(\tau_2))$.

\section{K3 fibrations}
\label{K3fibration}

\subsection{The work of Clingher-Doran}
Clingher and Doran introduced the following four-parameter quartic family in $\mathbb{P}^3$ \cite[Eq.~(3)]{arXiv:1004.3503} 
with canonical $\Lambda^{1,1} \oplus E_8(-1)\oplus E_7(-1)$ lattice polarization that generalizes a special two-parameter 
family of K3 surfaces introduced by Inose
\begin{equation}
\label{Inose}
\mathbf{Y}^2\mathbf{ZW}-4\, \mathbf{X}^3\mathbf{Z}+3\, \alpha \, \mathbf{XZW}^2 + \beta \, \mathbf{ZW}^3 + \gamma \, \mathbf{XZ}^2 \mathbf{W} -\frac{1}{2}(\delta \, 
\mathbf{Z}^2\mathbf{W}^2+\mathbf{W}^4)=0.
\end{equation}
They also find the parameters $(\alpha,\beta,\gamma,\delta)$ in terms of Siegel modular forms
\begin{equation}
 (\alpha,\beta,\gamma,\delta) = \left(\psi_4, \psi_6, 2^{12}3^5 \, \mathcal{C}_{10},
2^{12}3^6 \, \mathcal{C}_{12}\right) \;.
\end{equation}
(A similar picture was developed in earlier work for the case of a $H \oplus E_8 \oplus E_8$ lattice polarization \cite{math.AG/0602146}.)

Clingher and Doran determine an alternate elliptic fibration on (\ref{Inose}) that has two disjoint sections and a singular fiber of Kodaira-type $I_{10}^*$.
Here, we use a normalization consistent with F-theory and set
\begin{equation}
 \mathbf{X} = \dfrac{T \, X^3}{2^9 \, 3^5} \;, \quad \mathbf{Y}=\dfrac{X^2 \, Y}{2^{15/2} \, 3^{9/2}} \;,\quad \mathbf{W}=\dfrac{X^3}{2^{10} \, 3^6} \;, \quad \mathbf{Z}= \dfrac{X^2}{2^{16} \, 3^9} \;,
\end{equation}
and obtain from Equation~(\ref{Inose}) the Jacobian elliptic fibration
\begin{equation}
\label{WEq.bak.alt}
 Y^2 = X^3  + \left( T^3 - \frac{\psi_4}{48} \, T - \frac{\psi_6}{864} \right) \, X^2 -  \Big( 4 \, \mathcal{C}_{10}  \, T - \mathcal{C}_{12} \Big) \, X\;
\end{equation}
with special fibers of Kodaira-types $I_{10}^*$, $I_2$, and $6 \, I_1$, and the second section $(Y,X)=(0,0)$.
However, we are interested in the Jacobian elliptic fibration with two distinct special fibers of Kodaira-types $II^*$ and $III^*$, respectively. 
Therefore, we set
\begin{equation}
 \mathbf{X} = \dfrac{t \, x}{2^9 \, 3^5} \;, \quad \mathbf{Y}=\dfrac{y}{2^{15/2} \, 3^{9/2}} \;,\quad \mathbf{W}=\dfrac{t^3}{2^{10} \, 3^6} \;, \quad \mathbf{Z}= \dfrac{t^2}{2^{16} \, 3^9} \;,
\end{equation}
and obtain from Equation~(\ref{Inose}) the Jacobian elliptic fibration
\begin{equation}
\label{WEq.bak}
 y^2 = x^3  - t^3 \, \left( \frac{\psi_4}{48} \, t + 4 \, \mathcal{C}_{10} \right) \, x + t^5 \, \left( t^2 - \frac{\psi_6}{864} \, t + \mathcal{C}_{12} \right) \;.
\end{equation}
Clingher and Doran also state \cite[Thm.~1.7]{arXiv:1004.3503} that
\begin{equation}
\begin{split}
 \left[ I_2 : I_4 : I_6 : I_{10} \right]  = \left[ 2^3 \, 3 \, \delta : 2^2 3^2 \alpha \gamma^2, 2^3\, 3^2\, (4 \alpha \delta+ \beta\gamma)\gamma^2: 2^2 \gamma^6 \right] 
\end{split}
\end{equation}
which equals
\begin{equation}
\label{IgusaClebschProjective.bak}
\begin{split}
 \Big[ 2^3 \, 3 \, (3r\mathcal{C}_{12})\, : \, 2^2 3^2 \, \psi_4 \,  (r\mathcal{C}_{10})^2 \, : \, 2^3\, 3^2\, \Big(4 \psi_4  \, (3r\mathcal{C}_{12})+ \psi_6 \,(r\mathcal{C}_{10}) \Big)\, (r\mathcal{C}_{10})^2: 2^2 \,  (r\mathcal{C}_{10})^6 \Big]
 \end{split}
\end{equation}
with $r=2^{12} 3^5$.
Equation (\ref{IgusaClebschProjective.bak}) implies $\mathcal{C}_{10}=\chi_{10}$ and $\mathcal{C}_{12}=\chi_{12}$ by comparison with Equation (\ref{IgusaClebschProjective}).
This choice makes Equation (\ref{WEq.bak}) also be in agreement with Equation (\ref{eq:imp}).\footnote{Clingher and Doran claim $\mathcal{C}_{12}=\tilde{\chi}_{12}$ instead of $\mathcal{C}_{12}=\chi_{12}$, 
but we believe that this is the same slip as the one discussed in Equation~(\ref{definition2b})}

\subsection{The work of Kumar}
To relate this to Kumar's work, we must consider Igusa--Clebsch
invariants.  Kumar worked with the moduli space of curves of genus $2$,
which correspond to $\chi_{10}\ne0$.  Kumar's basic theorem \cite[Theorem~11]{MR2427457}  states that
a Weierstrass model for a family of K3 surfaces with $\Lambda^{1,1} \oplus E_8(-1)\oplus E_7(-1)$ lattice polarization is given
by the equation
\begin{equation}
\label{WE_MV2}
 y^2 = x^3 + t^3 \, (a \, t + c) \, x + t^5 \, (e \, t^2 + b \, t +d) \;,
\end{equation}
where $t$ is an affine coordinate on the base $\mathbb{P}^1$, $x, y$ are the affine coordinates on the fiber,
and the parameters $(a,b,c,d,e)$ are expressed in terms of the Igusa-Clebsch invariants from Section~\ref{moduli_curves_genus2} as follows:
\begin{equation}
 \begin{split}
 a & = - \dfrac{I_4}{12} = - \dfrac{\psi_4(\underline{\tau})}{2^6 \, 3^3} \;,\\
 b & = \dfrac{I_2 \, I_4 - 3 \, I_6}{108} = - \dfrac{\psi_6(\underline{\tau})}{2^8 \, 3^6}\;,\\
 c & = -1 \;,\\
 d & = \dfrac{I_2}{24} = \dfrac{\chi_{12}(\underline{\tau})}{2^3 \, 3 \, \chi_{10}(\underline{\tau})} \;,\\
 e & = \dfrac{I_{10}}{4} = \dfrac{\chi_{10}(\underline{\tau})}{2^3 \, 3^5}  \;.
 \end{split}
\end{equation}
Here, we used Equations (\ref{invariants}) to express the parameters in terms of Siegel modular forms.
The discriminant of the elliptic fiber in Equation (\ref{WE_MV2}) is
\begin{equation}\label{Delta_full}
\begin{split}
\Delta & =  t^9 \; 
\Big( 27 \, e^2 \, t^5+54 \, e \, b \, t^4+54 \, e \, d \, t^3+27 \, b^2 \, t^3 \\
& + 4 \, a^3 \, t^3+12 \, a^2 \, c \, t^2+54 \, b \, d \, t^2+12 \, a \,  c^2 \, t+27 \, d^2 \,t+4 \, c^3 \Big) \;.
\end{split}
\end{equation}
Generically, the fibration has a singular fiber of Kodaira-type $III^*$ at $t=0$ and a singular fiber of Kodaira-type $II^*$ at $t=\infty$.
Moreover, there are five singular fibers of Kodaira-type $I_1$ at those $t$ where the degree-five part of the discriminant vanishes. 
The Mordell-Weil group is $\mathrm{MW}(\pi)=\lbrace \mathrm{id} \rbrace$, the N\'eron-Severi lattice
has signature $(1,16)$ and discriminant $2$, and the transcendental lattice is $\Lambda^{2,2} \oplus \langle -2 \rangle$. 

We know from Section~\ref{Siegel_modular_forms} that $\chi_{10}(\underline{\tau}) \to 0$ as $z \to 0$ in $\underline{\tau}$. For the Weierstrass equation
to remain well-defined for $z \to0$, we rescale Equation~(\ref{WE_MV2})
as follows
\begin{equation}
 x \mapsto \dfrac{x}{\mu^6 \, \chi^2_{10}(\underline{\tau})} \;, \quad y \mapsto \dfrac{y}{\mu^9 \, \chi^3_{10}(\underline{\tau})} \;, \quad 
 t \mapsto \dfrac{9 \, t}{\chi_{10}(\underline{\tau})} 
\end{equation}
with $\mu = 2^{1/6}/3^{1/2}$. We then obtain the following rescaled parameters in Equations~(\ref{WE_MV2})
\begin{equation}
\label{duality_equation}
 a = -\dfrac{\psi_4(\underline{\tau})}{48} \;, \quad b = - \dfrac{ \psi_6(\underline{\tau}) }{864}\;, \quad
 c = - 4 \, \chi_{10}(\underline{\tau}) \;, \quad d=\chi_{12}(\underline{\tau}) \;, \quad e= 1\;.
\end{equation} 
With this choice for the coefficients Equation~(\ref{WE_MV2})
remains well-defined in the limit $z\to 0$. In fact, setting $z=0$ we obtain
\begin{equation}
 a = - 3 \, \dfrac{E_4(\tau_1) \, E_4(\tau_2)}{2^4 \, 3^2} \;, \quad b = - 2 \, \dfrac{ E_6(\tau_1) \, E_6(\tau_2)}{2^6 \, 3^3}\;, \quad
 c = 0 \;, \quad d=\eta(\tau_1)^{24} \, \eta(\tau_2)^{24} \;, \quad e= 1\;,
\end{equation} 
and after rescaling by $y\mapsto \lambda^{18} y$, $x\mapsto \lambda^{12} x$, $t\mapsto \lambda^6 \, t$ with $\lambda=\eta(\tau_1)^{2} \, \eta(\tau_2)^{2}$
the Weierstrass equation
 \begin{equation}
 \label{MV}
 y^2 = x^3  - 3 \,  A \, t^4 \, x + t^5 \, \big( t^2 - 2 \, B \, t + 1\big) \;,
\end{equation}
with
\begin{equation}
 A = \dfrac{E_4(\tau_1) \, E_4(\tau_2)}{2^4 \, 3^2 \, \eta(\tau_1)^{8} \, \eta(\tau_2)^{8}} \;, \quad B = \dfrac{E_6(\tau_1) \, E_6(\tau_2)}{2^6 \, 3^3 \, \eta(\tau_1)^{12} \, \eta(\tau_2)^{12}} \;. 
\end{equation}
Equation (\ref{MV}) matches precisely the family presented in \cite{FCY2}. Therefore, this computation provides yet another independent check of the normalization of $\chi_{12}$ in Equation~(\ref{relation2}). 

\section{Degenerations and five-branes}
\label{degs_and_branes}
In this section we consider certain degenerations of the multi-parameter family of K3 surfaces in Equation~(\ref{WE_MV2}).
As we have seen, the parameters $a, b, c, d$ can be interpreted as Siegel modular forms of even degree using Equation~(\ref{duality_equation}) or,
equivalently, as the Igusa-Clebsch invariants of a binary sextic using Equation~(\ref{invariants}). On the other hand, Namikawa and 
Ueno gave a geometrical classification of all (degenerate) fibers in pencils of curves of genus two in \cite{MR0369362}.
Given a family of curves of genus two over the complex line with affine coordinate $u \in \mathbb{C}$ which is smooth over $\mathbb{C} \backslash \lbrace 0 \rbrace$,
a multi-valued holomorphic map into the Siegel upper half plane of degree two, i.e., the period map, can be defined that determines the family uniquely. Moreover, 
there are three invariants called `monodromy', `modulus point', and `degree' which determine the singular fiber at $u=0$ uniquely.
To each singular fiber, which is labeled in a fashion similar to Kodaira's classification of singular fibers of elliptic surfaces, Namikawa and 
Ueno give a one-parameter family of genus-two curves with a singular fiber of each given type over $u=0$. 

We note that this work of Namikawa and Ueno provides an important class of
examples of degenerations in our situation, but cannot be complete.  This
is because they studied degenerations of genus two curves (with modular
group $\sptwoz$) rather than of binary sextics (with modular group
$O^+(L^{2,3})$).  Nevertheless, their work provides an interesting first
start at studying degenerations and the associated five-branes. 

From their list, we took all families of genus-two curves from \cite{MR0369362} that develop 
degenerations of type $III$, in particular, parabolic points of type $[3]$ with monodromy of infinite order over $u=0$. These families realize all singular 
fibers with modulus point
$\bigl(\begin{smallmatrix}
\tau_1&z\\ z&\infty
\end{smallmatrix} \bigr)$ for $z\not=0$ or $z=0$ and with $\tau_1 \not = \infty$. The families are listed in Table~\ref{degenerations} along with the 
Namikawa-Ueno type of the singular fiber over $u=0$ and the modulus point. For the families in the table we computed the Igusa-Clebsch invariants
as polynomials in $u$ and determined their asymptotic expansion as $u \to 0$. By means of Equation~(\ref{WE_MV2}), each of the families of genus-two curves
then determines a degenerating family of K3 surfaces as $u$ approaches zero.  The degeneration consists of two elliptic surfaces meeting along a rational curve.
In the last column of Table~\ref{degenerations} we list the Kodaira-types of the singular fibers of these two rational elliptic surfaces. Among these singular fibers, 
the stable models for the period points $\bigl(\begin{smallmatrix}
\tau_1&z\\ z&\infty
\end{smallmatrix} \bigr)$ or $\bigl(\begin{smallmatrix}
\tau_1&0\\ 0&\infty
\end{smallmatrix} \bigr)$ are given by families with degenerations of Namikawa-Ueno type $[I_{n-0-0}]$ or $[I_n-I_0-m]$ with $m, n >0$.

To determine one of the rational components in the degeneration limit, let $a, b, c, d$ be polynomials in $u$. If $c(u), d(u) \to 0$ as $u \to 0$, we obtain a degeneration 
\begin{equation}
  y^2 = x^3 + a(0) \,t^4 \, x + \big(t + b(0) \big)\, t^6 \;,
\end{equation}
i.e., a $(4,6,12)$-point at $u=0$.  Blowing up by setting $y=t^3y_1$, $x=t^2x_1$, $u=tu_1$, we obtain as the proper transform at $u_1=0$ the rational
elliptic surface 
\begin{equation}
  y_1^2 = x_1^3 + a(0)  \, x_1 +  \big(t + b(0) \big) 
\end{equation}
with singular fibers of Kodaira-types $II^* , 2 \, I_1$. This rational elliptic surface further degenerates to an isotrivial rational
elliptic surface with singular fibers $II^*, II$ and $j=0$ if $a(0)=0$. 

To determine the second rational component in the degeneration limit, we will have to consider different vanishing orders for the coefficients $a, b, c, d$ corresponding to different Namikawa-Ueno types for the singular fiber of the family of genus-two curves over $u=0$. 
As an example, we first consider the family of genus-two curves
\begin{equation}
 y^2 = u^\kappa \, \big(x^3+ \alpha x + 1\big) \, \big( (x-\beta)^2 + u^n \big)
\end{equation}
that develops a singularity of Namikawa-Ueno type $[I_{n-0-0}]$ and $[I^*_{n-0-0}]$ with modulus point
$\bigl(\begin{smallmatrix}
\tau_1&*\\ *&\infty
\end{smallmatrix} \bigr)$ for $\kappa=0$ and $\kappa=1$, respectively.
One checks that the asymptotic behavior of
the Igusa-Clebsch invariants for this family of genus-two curves
is given by
\begin{equation}
\begin{split}
 a(u)&=a_0 \, u^{4\kappa} + a_1 \, u^{4\kappa+n} + O(u^{4\kappa+2n})  \;,\\
 b(u)&=b_0 \, u^{6\kappa} + b_1 \, u^{6\kappa+n} + O(u^{6\kappa+2n}) \;,\\
 c(u)&= c_0 \, u^{10\kappa+n} + c_1 \, u^{10\kappa+2n} + O(u^{10\kappa+3n}) \;,\\
 d(u)&= d_0 \, u^{12\kappa+n} + d_1 \, u^{12\kappa+2n} + O(u^{12\kappa+3n}) \;,
\end{split}
\end{equation}
with $n>0$, $\kappa \in \lbrace 0,1 \rbrace$, and $a_0, b_0, c_0, d_0 \not=0$ and generic.
From our discussion above, it follows that the proper transforms in the coordinate chart $(u_1,t,x_1,y_1)$ at $u_1=0$ are
rational elliptic surfaces with singular fibers $II^*, 2\, I_1$ and $II^*, II$ for $\kappa=0$ and $\kappa=1$, respectively. 
On the other hand, setting $y=u^{21\kappa+3n} y_2$, $x=u^{14\kappa+2n} x_2$, $t=u^{6\kappa+n} t_2$ we obtain
the rational elliptic surface
\begin{equation}
  y_2^2 = x_2^3 + t_2^3 \, \big(a_0 t_2  + c_0 \big) \, x_2 +  t_2^5 \, \big(b_0 \, t_2 + d_0 \big) 
\end{equation}
with singular fibers $III^* , 3 \, I_1$ as the proper transform at $u=0$.

For each example in Table~\ref{degenerations}, we constructed the corresponding family of degenerating K3 surfaces and recovered the two 
rational elliptic surfaces in the degeneration limit whose singular fibers are listed in the last column of Table \ref{degenerations}.
We recorded the leading exponents $\mu(a), \mu(b), \mu(c), \mu(d)$ in the asymptotic expansions of $a, b, c, d$
in Table \ref{degenerations2} where $\kappa=1$ if there is an additional star-fiber and $\kappa=0$ otherwise.
In Table \ref{degenerations2}, we also recorded the exponents $\mu(y), \mu(x), \mu(u)$ used in the coordinate change $y=u^{\mu(y)} y_2$, 
$x=u^{\mu(x)} x_2$, $t=u^{\mu(t)} t_2$ that recovers the second rational component in the degeneration limit.

\bigskip

\begin{table}
{\footnotesize
\begin{tabular}{|c|c|c|c|l|}
type & modulus & family of genus-two curves & rat. components \\
\hline
\hline
&&&\\[-0.9em]
$[I_{n-0-0}]$ & $\bigl(\begin{smallmatrix}
\tau_1&*\\ *&\infty
\end{smallmatrix} \bigr)$ &
$y^2 = \big(x^3+ \alpha x + 1\big) \, \big( (x-\beta)^2 + u^n \big)$ &
$\begin{array}{c} II^*, 2 I_1\\ III^*, 3 I_1\end{array}$\\[0.4em]
\hline
&&&\\[-0.9em]
$[I^*_{n-0-0}]$ & $\bigl(\begin{smallmatrix}
\tau_1&*\\ *&\infty
\end{smallmatrix} \bigr)$ &
$y^2 = u \, \big(x^3+ \alpha x + 1\big) \, \big( (x-\beta)^2 + u^n \big)$ &
$\begin{array}{c} II^*, II\\ III^*, 3 I_1\end{array}$\\[0.4em]
\hline
&&&\\[-0.9em]
$[II_{n-0}]$ & $\bigl(\begin{smallmatrix}
\tau_1&*\\ *&\infty
\end{smallmatrix} \bigr)$ &
$y^2 =  \big(x^4+ \alpha \, u \, x^2 + u^2\big) \, \big( (x-1)^2 + u^{n-1} \big)$ &
$\begin{array}{c} II^*, II\\ III^*, III\end{array}$\\[0.4em]
\hline
&&&\\[-0.9em]
$[II^*_{n-0}]$ & $\bigl(\begin{smallmatrix}
\tau_1&*\\ *&\infty
\end{smallmatrix} \bigr)$ &
$y^2 =  u \, \big(x^4+ \alpha \, u \, x^2 + u^2\big) \, \big( (x-1)^2 + u^{n-1} \big)$ &
$\begin{array}{c} II^*, II\\III^*, III\end{array}$\\[0.4em]
\hline
\hline
&&&\\[-0.9em]
$[I_n-I_0-m]$ & $\bigl(\begin{smallmatrix}
\tau_1&0\\ 0&\infty
\end{smallmatrix} \bigr)$ & $y^2 = \big(x^3+ \alpha \, u^{4m} \, x + u^{6m}\big) \, \big( (x-1)^2 + u^n \big)$ &
$\begin{array}{c} II^*, II\\ II^*, 2 I_1 \end{array}$\\[0.4em]
\hline
&&&\\[-0.9em]
$[I_n-I^*_0-m]$ & $\bigl(\begin{smallmatrix}
\tau_1&0\\ 0&\infty
\end{smallmatrix} \bigr)$ & $y^2 = \big(x^3+ \alpha \, u^{4m+2} \, x + u^{6m+3}\big) \, \big( (x-1)^2 + u^n \big)$ &
$\begin{array}{c} II^*, II\\ II^*, II \end{array}$\\[0.4em]
\hline
&&&\\[-0.9em]
$[I_n-I^*_0-0]$ & $\bigl(\begin{smallmatrix}
\tau_1&0\\ 0&\infty
\end{smallmatrix} \bigr)$ & $y^2 = \big(x^3+ \alpha \, u^{2} \, x + u^{3}\big) \, \big( (x-1)^2 + u^n \big)$ &
$\begin{array}{c} II^*, II\\ III^*, II, I_1 \end{array}$\\[0.4em]
\hline
&&&\\[-0.9em]
$[I_0-I_n^*-m]$ & $\bigl(\begin{smallmatrix}
\tau_1&0\\ 0&\infty
\end{smallmatrix} \bigr)$ & $\begin{array}{c}y^2 = \big(x+u\big) \, \big(x^2+u^{n+2}\big) \, \\[0.2em] \times \big((x-1)^3+ \alpha \, u^{4m} \, (x-1) + u^{6m}\big)\end{array}$ &
$\begin{array}{c} II^*, II\\II^*, II\end{array}$\\[0.6em]
\hline
&&&\\[-0.9em]
$[I_0-I_n^*-0]$ & $\bigl(\begin{smallmatrix}
\tau_1&0\\ 0&\infty
\end{smallmatrix} \bigr)$ & $\begin{array}{c}y^2 = \big(x+u\big) \, \big(x^2+u^{n+2}\big) \,\\[0.2em] \times  \big((x-1)^3+ \alpha \,  (x-1) + 1\big)\end{array}$ &
$\begin{array}{c} II^*, II\\III^*, II, I_1\end{array}$\\[0.6em]
\hline
&&&\\[-0.9em]
$[I^*_0-I_n^*-m]$ & $\bigl(\begin{smallmatrix}
\tau_1&0\\ 0&\infty
\end{smallmatrix} \bigr)$ & $\begin{array}{c}y^2 = \big(x+u\big) \, \big(x^2+u^{n+2}\big) \,\\[0.2em] \times \, \big((x-1)^3+ \alpha \, u^{4m+2} \, (x-1) + u^{6m+3}\big)\end{array}$ &
$\begin{array}{c} II^*, II\\II^*, II\end{array}$\\[0.6em]
\hline
&&&\\[-0.9em]
$[I^*_0-I_n^*-0]$ & $\bigl(\begin{smallmatrix}
\tau_1&0\\ 0&\infty
\end{smallmatrix} \bigr)$ & $\begin{array}{c} y^2 = \big(x+u\big) \, \big(x^2+u^{n+2}\big) \,\\[0.2em] \times \big((x-1)^3+ \alpha \, u^{2} \, (x-1) + u^{3}\big)\end{array}$ &
$\begin{array}{c} II^*, II\\III^*, II, I_1\end{array}$\\[0.6em]
\hline
\end{tabular}}
\medskip
\caption{Families of genus-two curves with degeneration of type $III$}\label{degenerations}
\end{table}

\begin{table}
\begin{center}
\scalebox{0.88}{
{\footnotesize
\begin{tabular}{|c||c|c|c|c|c|c|c|}
type & $\mu(a)$ & $\mu(b)$ & $\mu(c)$ & $\mu(d)$ & $\mu(y)$ & $\mu(x)$ & $\mu(t)$ \\
\hline
\hline
&&&&&&&\\[-0.8em]
$\begin{array}{c} [I_{n-0-0}]\\[0.2em] [I^*_{n-0-0}] \end{array}$ 
& $4\kappa$ & $6\kappa$ & $10\kappa+n$ & $12\kappa+n$ & $21\kappa+3n$ & $14\kappa+2n$ & $6\kappa+n$ \\[1em]
\hline
&&&&&&&\\[-0.8em]
$\begin{array}{c} [II_{n-0}]\\[0.2em] [II^*_{n-0}]\end{array}$ 
& $2+4\kappa$ & $3+6\kappa$ & $\begin{array}{c}5\,+10\kappa\\+\,n\end{array}$ 
& $\begin{array}{c}6\,+12\kappa\\+\,n\end{array}$ & $\begin{array}{c}15+21\kappa\\ \; \; \, +\, 3n\end{array}$ & $\begin{array}{c}10\; +14\kappa\\\quad +\, 2n\end{array}$ & $\begin{array}{c}5+6\kappa\\\;+ \,n\end{array}$ \\[1em]
\hline
&&&&&&&\\[-0.8em]
$\begin{array}{c} [I_{n}-I_0-m]\\[0.2em] [I_n-I_0^*-m] \end{array}$ 
& $2\kappa+4m$ & $3\kappa+6m$ & $\begin{array}{c}6\kappa+12m\\+\,n\end{array}$ 
& $\begin{array}{c}6\kappa+12m\\ +\,n\end{array}$ & $\begin{array}{c}18\kappa+21m\\\;\;\;\,+\, 3n\end{array}$ & $\begin{array}{c}12\kappa+14m\\\;\;\;\,+\, 2n\end{array}$ & $\begin{array}{c}6\kappa+6m\\\;\;+ \,n\end{array}$ \\[1em]
\hline
&&&&&&&\\[-0.8em]
$\begin{array}{c} [I_{0}-I_n^*-m]\\[0.2em] [I_0^*-I_n^*-m] \end{array}$ 
& $\begin{array}{c}2\kappa+4m\\\;\,+\,2\end{array}$ & $\begin{array}{c}3\kappa+6m\\\;\,+\, 3\end{array}$ & $\begin{array}{c}6\kappa+12m\\+\,6+n\end{array}$ 
& $\begin{array}{c}6\kappa+12m\\+\,6+n\end{array}$ & $\begin{array}{c}18\kappa+21m\\+\, 18+ 3n\end{array}$ & $\begin{array}{c}12\kappa+14m\\+\, 12+ 2n\end{array}$ & $\begin{array}{c}6\kappa+6m\\+ \,6+n\end{array}$ \\[1em]
\hline
\end{tabular}}}
\end{center}
\medskip
\caption{Exponents of coefficients for families of K3 surfaces}\label{degenerations2}
\end{table}

\end{document}